\documentclass[review,onefignum,onetabnum]{article}

\usepackage{latexsym}
\usepackage{amssymb,amsbsy,amsmath,amsfonts,amssymb,amscd,amsthm}
\usepackage{subfigure}
\usepackage{graphicx}
\usepackage{hyperref}
\usepackage{dsfont}
\usepackage{mathtools}
\usepackage{pdfsync}
\usepackage{epsfig,epstopdf}
\usepackage{caption}
\usepackage{xcolor}
\usepackage{hyperref}

\usepackage{amsmath,mathrsfs,amsfonts,upgreek,graphicx,mathtools,nicefrac,color,dsfont}

\newtheorem{ass}[]{Assumption}

\newcommand{\beq}{\begin{equation}}
\newcommand{\eeq}{\end{equation}}
\newcommand{\be}{\begin{equation}}
\newcommand{\ee}{\end{equation}}

\usepackage{stmaryrd} % enable llb, rrb

\newcommand*\diff{\mathop{}\!\mathrm{d}}

\newcommand {\e}  {\varepsilon}

\numberwithin{equation}{section}
\numberwithin{prpstn}{section}
\numberwithin{rmrk}{section}
\newtheorem{lemma}{Lemma}
\newtheorem{definition}{Definition}
\newtheorem{rem}{Remark}

\title{Macroscopic limit of a kinetic model describing the switch in T cell migration modes via binary interactions}

\author{Gissell Estrada-Rodriguez\thanks{Sorbonne Universit\'e, CNRS, Universit\'e de Paris, Inria, Laboratoire Jacques-Louis Lions UMR7598, F-75005 Paris, France (estradarodriguez@ljll.math.upmc.fr)}
\and
Tommaso Lorenzi\thanks{Department of Mathematical Sciences ``G. L. Lagrange'', Dipartimento di Eccellenza 2018-2022, Politecnico di Torino, 10129 Torino, Italy (tommaso.lorenzi@polito.it)}
}

\begin{document}
\date{}
\maketitle

\begin{abstract}
Experimental results on the immune response to cancer indicate that activation of cytotoxic T lymphocytes (CTLs) through interactions with dendritic cells (DCs) can trigger a change in CTL migration patterns. In particular, while CTLs in the pre-activation state move in a non-local search pattern, the search pattern of activated CTLs is more localised. In this paper, we develop a kinetic model for such a switch in CTL migration modes. The model is formulated as a coupled system of balance equations for the one-particle distribution functions of CTLs in the pre-activation state, activated CTLs and DCs. CTL activation is modelled via binary interactions between CTLs in the pre-activation state and DCs. Moreover, cell motion is represented as a velocity-jump process, with the running time of CTLs in the pre-activation state following a long-tailed distribution, which is consistent with a L\'{e}vy walk, and the running time of activated CTLs following a Poisson distribution, which corresponds to Brownian motion. We formally show that the macroscopic limit of the model comprises a coupled system of balance equations for the cell densities whereby activated CTL movement is described via a classical diffusion term, whilst a fractional diffusion term describes the movement of CTLs in the pre-activation state. The modelling approach presented here and its possible generalisations are expected to find applications in the study of the immune response to cancer and in other biological contexts in which switch from non-local to localised migration patterns occurs.
\end{abstract}

% REQUIRED
%\begin{keywords}
%continuum mechanics; thin membranes; effective interface conditions; cell invasion; basement membrane; ovarian cancer
%\end{keywords}

% REQUIRED
%\begin{AMS}
%35Q92; 35R05; 92C10; 92C17
%\end{AMS}

\section{Introduction}
The interaction between dendritic cells (DCs) and cytotoxic T lymphocytes (CTLs) plays a pivotal role in the immune response to cancer. DCs recognise the antigens expressed by cancer cells and present them to CTLs, which then become selectively activated against those antigens~\cite{waldman2020guide,wculek2019dendritic}. Growing experimental evidence indicates that activation of CTLs via antigen presentation by DCs can bring about a switch in CTL migration modes~\cite{boissonnas2007vivo,krummel2016t}. In fact, while CTLs in the pre-activation state move in a non-local search pattern, which enables them to rapidly scan DCs for the presence of possible tumour antigens, the search pattern of activated CTLs is more localised. This allows activated CTLs to stay within a confined area for longer, thus facilitating their encounter with tumour cells expressing the antigens they have been activated against.

Stochastic individual-based models of immune response to cancer taking explicitly into account this difference in movement between CTLs have recently been developed~\cite{macfarlane2019stochastic,macfarlane2018modelling}. In these models, cell motion is described as a space-jump process~\cite{othmer1988models}. In particular, CTLs in the pre-activation state undergo a space-jump process consistent with a L\'{e}vy walk, whereas a space-jump process corresponding to Brownian motion is used to describe the movement of activated CTLs. Such individual-based models enable representation of biological processes at the level of single cells and account for possible stochastic variability in cell dynamics, which allow for greater adaptability and higher accuracy in mathematical modelling. However, as the numerical exploration of these models requires large computational times for clinically relevant cell numbers (e.g. cell numbers of orders of magnitude between $10^6$ and $10^9$~\cite{azizi2018single}) and the models are not analytically tractable, it is desirable to derive corresponding deterministic continuum models in a suitable limit.

In this paper, integrating the ideas proposed in~\cite{macfarlane2019stochastic,macfarlane2018modelling} with the modelling approach presented in~\cite{estrada2020interacting,estrada2018fractional}, we develop a kinetic model for the switch in CTL migration modes that is caused by activation through interactions with DCs. Cells are grouped into three populations: CTLs in the pre-activation state (i.e. inactive CTLs), activated CTLs and DCs. In the model, DCs are assumed to present a given tumour antigen on their surface so that they can activate inactive CTLs by contact. Since the focus of this study is on the mathematical modelling of the change in CTL migration mode upon activation, we do not take into account biological processes involving cell division and death. Furthermore, for simplicity, we do not consider the occurrence of molecular processes leading activated CTLs to re-enter a pre-activation state~\cite{wherry2015molecular}.

The model is formulated as a coupled system of balance equations for the one-particle distribution functions of the three cell populations. CTL activation is modelled as a process of population switching among CTLs induced by binary interactions between inactive CTLs and DCs. Moreover, cell motion is represented as a velocity-jump process~\cite{othmer1988models}, with the running time of inactive CTLs following a long-tailed distribution, which is consistent with a L\'{e}vy walk~\cite{estrada2020interacting,estrada2018fractional}, and the running time of activated CTLs following a Poisson distribution, which corresponds to Brownian motion. Using a method similar to that previously employed in~\cite{estrada2020interacting}, we formally show that the macroscopic limit of this model comprises a coupled system of balance equations for the cell densities, whereby activated CTL movement is described via a classical diffusion term, whilst a fractional diffusion term describes the movement of CTLs in the pre-activation state.  

The paper is organised as follows. In Section~\ref{Sec:micromodel}, we introduce the modelling strategies and the main assumptions used to describe the spatio-temporal dynamics of CTLs and DCs at the scale of single cells, which provide a microscopic representation of the biological system. In Section~\ref{Sec:mesomodel}, we present the kinetic model, which constitutes a mesoscopic analogue of the underlying microscopic scale model. In Section~\ref{Sec:macromodel}, we derive the macroscopic limit of a suitably rescaled version of the kinetic model. Section~\ref{Sec:persp} concludes the paper providing a brief overview of possible research perspectives.

\section{Description of the system at the microscopic scale}
\label{Sec:micromodel}
\paragraph{Biological system and cell populations} We label the three cell populations by a letter $h \in \{A,D,I\}$, that is, activated CTLs are labelled by $h=A$, DCs are labelled by $h=D$ and inactive CTLs are labelled by $h=I$. We let the total number of cells in the system be denoted by $N = N_D + N_T$, where $N_D \in \mathbb{N}$ is the number of DCs and $N_T \in \mathbb{N}$ is the total number of CTLs. Moreover, we describe the number of inactive and activated CTLs in the system at time $t \in \mathbb{R}_+$ by means of the functions $N_I(t)$ and $N_A(t)$, respectively, with $N_I(t) + N_A(t) = N_T$ for all $t$. 

\paragraph{Mathematical representation of individual cells} Every individual cell is modelled as a sphere of diameter $\varrho \in \mathbb{R}^*_+$ and is labelled by an index $i = 1, \ldots, N$. The phase-space state of the $i^{th}$ cell is represented by a pair $(\mathbf{x}_i, \mathbf{v}_i)$, where the vector $\mathbf{x}_i\in \mathbb{R}^n$ describes the position of the centre of the cell and the vector $\mathbf{v}_i\in {\rm V} \subset \mathbb{R}^n$, with ${\rm V} :=\{\mathbf{v}_i\in\mathbb{R}^n:\ |\mathbf{v}_i|=1 \}$ (i.e. ${\rm V}$ is the unit $n$-sphere), represents the direction of the cell velocity. Moreover, the magnitude of the cell velocity is assumed to be constant and is denoted by $c \in \mathbb{R}^*_+$. The value of $n=1,2,3$ depends on the biological scenario under study.

\subsection{Description of cell motion}
\label{Sec:micromodelmot}
\paragraph{Velocity-jump process} We describe the motion of a cell labelled by an index $i$ as a run-and-tumble process with run time $\tau_i \in \mathbb{R}^*_+$ and running probability $\psi(\mathbf{x}_i,\tau_i)$, where $0< \psi(\cdot,\cdot) \leq 1$ and $\partial_{\tau_i}\psi(\cdot,\cdot) \leq 0$. The running probability $\psi(\mathbf{x}_i,\tau_i)$ correlates with the stopping rate $\beta(\mathbf{x}_i,\tau_i)$ through the relations given by the following definition~\cite{estrada2020interacting}
\begin{equation}
\psi(\mathbf{x}_i,\tau_i)\coloneqq \exp\left(\int_0^{\tau_i}\beta(\mathbf{x}_i,s)\diff s \right) \ , \quad \beta = \frac{\varphi}{\psi} \quad \text{with} \quad \varphi :=  -\partial_{\tau_i}\psi \ .\label{eq: beta} 
\end{equation}
Hence, starting at position $\mathbf{x}_i$ at time $t$, the $i^{th}$ cell will continue moving along a straight path in the direction given by the vector $\mathbf{v}_i$ with constant speed $c$ for a period of time $\tau_i$, after which it may stop with rate $\beta(\mathbf{x}_i,\tau_i)$. The cell will then instantaneously resume moving in a new randomly selected direction given by a vector $\bar{\mathbf{v}}_i$, which is prescribed by a turning kernel $\ell(\mathbf{x}_i,t,\mathbf{v}_i;\bar{\mathbf{v}}_i)$ -- i.e. cells undergo a velocity-jump process~\cite{othmer1988models}.

\paragraph{Running probability} 
The running probability $\psi(\mathbf{x}_i,\tau_i)$ determines the distribution of the running time $\tau_i$ and depends on the way in which the $i^{th}$ cell moves. Note that the running probability is here assumed to be independent from the cell velocity $\mathbf{v}_i$. On the basis of experimental evidence reported in~\cite{boissonnas2007vivo,engelhardt2012marginating}, we assume that inactive CTLs move in a non-local search pattern corresponding to trajectories that are characterised by a strong presence of long runs, which enable them to cover larger areas. On the other hand, activated CTLs and DCs\footnote{We remind the reader that we consider DCs presenting a given tumour antigen on their surface.} move in a more localised search pattern. In particular, building upon the modelling approach presented in~\cite{macfarlane2018modelling}, we describe the motion of activated CTLs and DCs as a Brownian motion, whereas we let inactive CTLs undergo superdiffusive motion consistent with a L\'{e}vy walk, whereby the mean square displacement grows nonlinearly with time. In particular, the mean-square displacement at time $t$ is proportional to $t^{\nicefrac{2}{\alpha}}$, where $\alpha \in (1,2)$ is the L\'{e}vy exponent. We recall that $\alpha=1$ and $\alpha=2$ would correspond to ballistic motion (i.e. a form of motion whereby the mean-square displacement at time $t$ is proportional to $t^2$) and classical diffusion, respectively.

Under these assumptions, if the $i^{th}$ cell belongs to population $A$ or population $D$, we let the value of the running time $\tau_i$ follow a Poisson distribution~\cite{othmer2000diffusion}. Hence, under the additional simplifying assumption that cells in populations $A$ and $D$ are characterised by the same stopping rate, which is assumed to be constant and thus modelled by a parameter $b  \in \mathbb{R}^*_+$, we use the following definition of the running probability
\begin{equation}
\psi(\mathbf{x}_i,\tau_i) \equiv \psi(\tau_i) := \exp\left(-b \, \tau_i \right), \quad \varphi(\mathbf{x}_i,\tau_i) \equiv \varphi(\tau_i) := b \, \exp\left(-b \, \tau_i \right) \ .
\label{eq: brownian}
\end{equation}
On the other hand, if the $i^{th}$ cell belongs to population $I$, we let the value of the running time $\tau_i$ follow a long-tailed distribution, and we define the running probability along the lines of~\cite{estrada2020interacting} as
\begin{equation}
\psi(\mathbf{x}_i,\tau_i):=\Bigl(\frac{\tau_0(\mathbf{x}_i)}{\tau_0(\mathbf{x}_i)+\tau_i} \Bigr)^\alpha\ , \quad \varphi(\mathbf{x}_i,\tau_i):=\frac{\alpha \, \tau_0(\mathbf{x}_i)^\alpha}{(\tau_0(\mathbf{x}_i)+\tau_i)^{\alpha+1}}, \quad \alpha\in(1,2) \ .\label{eq: levy}
\end{equation}
Here, the function $\tau_0(\mathbf{x}_i) \geq 0$ captures possible spatial inhomogeneities in the running time distribution.

\paragraph{Turning kernel and turning operator} We consider the case where the new direction of cell motion given by $\bar{\mathbf{v}}_i$ is symmetrically distributed with respect to the original direction given by $\mathbf{v}_i$ and, therefore, we let the turning kernel $\ell(\mathbf{x}_i,t,\mathbf{v}_i;\bar{\mathbf{v}}_i)$ satisfy the following assumptions~\cite{alt1980biased}
\begin{equation}
\label{ass:ell}
\ell(\mathbf{x}_i,t,\mathbf{v}_i;\bar{\mathbf{v}}_i) \equiv \ell(\mathbf{x}_i,t,|\bar{\mathbf{v}}_i-\mathbf{v}_i|) \ , \quad  \int_{\rm V} \ell(\cdot,\cdot,|\mathbf{v}_i-\mathbf{e}_1|)\diff \mathbf{v}_i=1 \ ,
\end{equation}
where $\mathbf{e}_1 = (1, 0, \ldots, 0) \in \mathbb{R}^n$ is a unit vector.

Moreover, we let the integral operator $\mathcal{T}$ be a turning operator such that for all test functions $\phi(\mathbf{v}_i)$
\begin{equation}
    \mathcal{T}[\phi](\cdot, \cdot, \bar{\mathbf{v}}_i) =\int_{\rm V} \ell(\cdot,\cdot,\mathbf{v}_i;\bar{\mathbf{v}}_i) \phi(\mathbf{v}_i) \diff \mathbf{v}_i \ ,  \label{eq: turn angle operator}
\end{equation}
where $\ell$ is the turning kernel defined via~\eqref{ass:ell}. Since $\displaystyle{\int_{\rm V} \ell(\cdot,\cdot,\cdot;\bar{\mathbf{v}}_i) \diff \bar{\mathbf{v}}_i =1}$, we have
\begin{equation}
    \int_{\rm V}(\mathds{1}-\mathcal{T})[\phi](\cdot, \cdot,\bar{\mathbf{v}}_i) \diff\bar{\mathbf{v}}_i=0\ ,\label{eq: zero turn operator}
\end{equation}
where $\mathds{1}$ is the identity operator. 

Finally, we recall that in $n$-dimensions the surface area of the unit sphere ${\rm V} $ is
\begin{equation}
|{\rm V}|=\begin{cases}
\dfrac{2\pi^{\nicefrac{n}{2}}}{\Gamma\left(\frac{n}{2}\right)} \ , & \text{for $n$ even},\\
\dfrac{\pi^{\nicefrac{n}{2}}}{\Gamma\left(\frac{n}{2}+1\right)} \ , & \text{for $n$ odd}
\end{cases}
\label{volumen}
\end{equation}
where $\Gamma(\cdot)$ is the Gamma function, and we also recall some useful properties of the spectrum of the turning operator $\mathcal{T}$~\cite{alt1980biased}:

\begin{lemma}\label{lem: eigenfunctions}
If the turning kernel $\ell(\cdot,\cdot,|\bar{\mathbf{v}}_i-\mathbf{v}_i|)$ is continuous, then $\mathcal{T}$ is a symmetric compact operator. In particular, there exists an orthonormal basis of $L^2({\rm V})$ consisting of eigenfunctions $\{ \phi_k,\ k\geq 0\}$ of $\mathcal{T}$. Using the notation $\mathbf{v}_i=({v}_0^i,{v}^i_1,\dots,{v}^i_{n-1}) \in {\rm V}$, we have:
$$
\begin{aligned}\phi_{0}(\mathbf{v}_i) & =\frac{1}{|{\rm V}|} &  & \text{is an eigenfunction associated with the eigenvalue} & & \iota_{0}=1,\\
\phi_{1}^j(\mathbf{v}_i) & =\frac{n{v}^i_j}{|{\rm V}|} &  & \text{are eigenfunctions associated with the eigenvalue} & & 
\end{aligned}
$$
\begin{equation}
\iota_{1}=\int_{\rm V}\ell(\cdot,\cdot,|\bar{\mathbf{v}}_i-\mathbf{e}|)\bar{v}^{i}_1\diff\bar{\mathbf{v}}_i<1,
\label{eq: eigen}
\end{equation}
where $\mathbf{e} = (1, 1, \ldots, 1) \in \mathbb{R}^n$ is the vector with all components equal to $1$. Moreover, any function $p_i\in L^2(\mathds{R}^n\times \mathds{R}_+\times {\rm V})$ admits a unique decomposition of the form
\begin{equation}
p_i=\frac{1}{|{\rm V}|}\left(\rho_i+n\mathbf{v}_i\cdot {w}_i \right)+\hat{z},\label{eq: real_eigen}
\end{equation}
where $\hat{z}$ is orthogonal to all linear polynomials in $\mathbf{v}_i$,
\begin{equation*}
\rho_i(\mathbf{x}_i,t)=\int_{\rm V}p_i(\mathbf{x}_i,t,\mathbf{v}_i)\phi_0(\mathbf{v}_i) \diff\mathbf{v}_i,\quad {w}_i^j(\mathbf{x}_i,t)=\int_{\rm V} p_i(\mathbf{x}_i,t,\mathbf{v}_i)\phi_1^j(\mathbf{v}_i) \diff\mathbf{v}_i,
\end{equation*}
and  ${w}_i =  ({w}^i_0, \dots, {w}^i_{n-1})$.
\end{lemma}

\subsection{Description of the interactions between cells}
\label{Sec:micromodelint}

\begin{figure}
    \centering
    \includegraphics[scale=0.4]{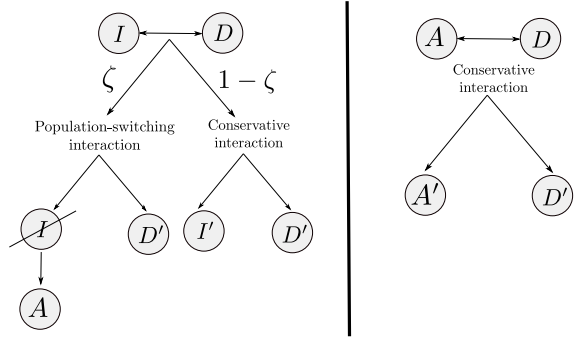}
    \caption{Schematics of cell-cell interactions corresponding to Assumptions~\ref{Ass1} and~\ref{Ass2}. Prime symbols indicate a change in cell velocity upon interaction.}\label{fig: collisions}
\end{figure}
Building on previous work on individual-based models of interaction dynamics between DCs and CTLs~\cite{macfarlane2019stochastic,macfarlane2018modelling}, we consider only the effects of binary cell-cell interactions, thus neglecting interactions that involve more than two cells. 

Moreover, given that the focus of this work is on modelling the switch in T cell migration modes mediated by interactions between inactive CTLs and DCs, we explicitly model the effects of interactions between cells of population $I$ and cells of population $D$, while for simplicity, we neglect the effects of intrapopulation cell-cell interactions and interactions between cells of population $I$ and cells of population $A$. 

Furthermore, the spatial dynamics of DCs are primarily affected by interactions with inactive CTLs~\cite{bousso2008t, gardner2016dendritic, rothoeft2006structure}. Hence, for simplicity, we explicitly model the effect of interactions between cells of population $D$ and cells of population $A$ on the motion of $A$ cells, while we neglect the effect of these interactions on the motion of $D$ cells, since we take it to be negligible compared to that of interactions with cells of population $I$.

On the basis of these considerations, we incorporate into the model only the effects of  interactions between {pairs of cells} that are summarised by the schematics in Figure~\ref{fig: collisions}, which correspond to the following definitions and assumptions. 

\begin{definition}{{\bf (Conservative interactions)}}
\label{Ass0}
Conservative interactions are those that preserve the number of cells in every population and  only modify the velocity of the cells according to \eqref{eq: new velocity}. Otherwise, the interaction is a population-switching interaction.
\end{definition}

\begin{definition}{\bf (Population-switching interactions)}
\label{Ass0b}
Population-switching interactions are those that lead a cell to enter a different population. These interactions are destructive for the  original population of the cell and creative for the population in which the cell will be upon interaction.
\end{definition}

\begin{ass}[{\bf Interactions between inactive CTLs and DCs}]
\label{Ass1}
\emph{We model activation of CTLs upon interaction with DCs by assuming that, when a cell in population $I$ interacts with a cell in population $D$, the $I$ cell switches from population $I$ to population $A$ (i.e. the interaction is population-switching in the sense of Definition~\ref{Ass0b}) with probability $\zeta \in (0,1)$. For simplicity, we assume that the $I$ cell enters population $A$ without changing its velocity. If activation does not occur, event that happens with probability $1-\zeta$, the $I$ cell remains in the same population (i.e. the interaction is conservative in the sense of Definition~\ref{Ass0}) and acquires the post-interaction velocity defined via~\eqref{eq: new velocity}. Upon interaction, the $D$ cell  always acquires a post-interaction velocity defined as in~\eqref{eq: new velocity}.} 
\end{ass}
\begin{ass}[{\bf Interactions between activated CTLs and DCs}]
\label{Ass2}
\emph{We assume that when a cell in population $A$ interacts with a cell in population $D$, the { $A$} cell remains in the same population and acquires the post-interaction velocity defined via~\eqref{eq: new velocity}, and the interaction is conservative in the sense of Definition~\ref{Ass0}. As explained above, we do not take into account the effect of interactions between cells of population $D$ and cells of population $A$  { on} the motion of { the $D$} cells.}
\end{ass}

\begin{figure}
    \centering
    \includegraphics[scale=1.3]{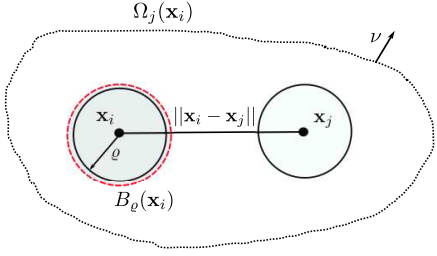}
    \caption{Schematics of the interaction domain defined in \eqref{Omegaxi}.}
    \label{domain schematics}
\end{figure}
\iffalse
\begin{ass}[{\bf Interactions between DCs and CTLs}]
\label{Ass3}
{ \emph{We assume that if a cell in population $D$ interacts with a cell in population $I$, the cell $D$ will acquire the post-\textcolor{magenta}{interaction} velocity defined via~\eqref{eq: new velocity} and the interaction will be conservative.  For simplicity, we do not take into account the effect of interactions between cells of population $D$ and cells of population $A$.}} 
\end{ass}
\fi
\iffalse
\begin{ass}[{\bf Other cell-cell interactions}]
\label{Ass4}
\emph{Since here we are primarily interested in the mathematical modelling of the switch in T cell migration modes mediated by interactions between inactive CTLs and DCs, for simplicity we neglect interactions between cells of the same population and we do not take into account the effect of interactions between cells of population $I$ and cells of population $A$.}
\end{ass}
\fi
We allow interactions between a cell {$i$} in the phase-space state $(\mathbf{x}_i, \mathbf{v}_i)$ and a cell {$j$} in the phase-space state $(\mathbf{x}_j, \mathbf{v}_j)$ to occur when the  cell {$j$} is in the domain of interaction of the cell {$i$}, which is defined as the set 
\begin{equation}
\label{Omegaxi}
\Omega_j(\mathbf{x}_i) := \{ \mathbf{x}_j\in\mathbb{R}^n: |\mathbf{x}_i-\mathbf{x}_j|\geq\varrho\} \equiv \mathbb{R}^n \setminus {\rm B}_\varrho(\mathbf{x}_i) \ ,
\end{equation}
where ${\rm B}_\varrho(\mathbf{x}_i)$ denotes the ball of radius $\varrho$ centred at $\mathbf{x}_i$. If a cell {$i$} acquires a new velocity upon interaction with a cell {$j$}, the new velocity is defined, for simplicity, as the following post-interaction velocity
\begin{equation}
\mathbf{v}'_i=\mathbf{v}_i - 2 \, (\mathbf{v}_i\cdot \nu) \, \nu \quad \text{with} \quad  \nu := \frac{\mathbf{x}_i-\mathbf{x}_j}{|\mathbf{x}_i-\mathbf{x}_j|}\label{eq: new velocity} \ ,
\end{equation}
where $\nu$ is the normal vector at the point of interaction (i.e. $\nu$ is the unit normal that points outward from $\Omega_j(\mathbf{x}_i)$ and inward to ${\rm B}_\varrho(\mathbf{x}_i)$)~\cite{cercignani2013mathematical}.

\begin{rem}
\label{rem:elast}
Definition~\eqref{eq: new velocity} relies on the observation that, although binary collisions between cells are not elastic in nature, they may result in cell outgoing trajectories compatible with those observed in elastic collisions~\cite{albrecht1977phagokinetic, lober2015collisions}.
\end{rem}

%In particular, we consider the binary interactions between \textcolor{magenta}{pairs of cells} that are summarised by the schematics in Figure~\ref{fig: collisions}, which correspond to the following assumptions. \textcolor{magenta}{Cell interactions
%may either modify the state of each cell or the number of cells in each population
%by destruction phenomena.}

\section{Mesoscopic scale model}
\label{Sec:mesomodel}
In this section, we derive the mesoscopic scale model corresponding to the microscopic scale description presented in Section~\ref{Sec:micromodel}, which comprises a system of transport equations for the one-particle distribution functions of inactive CTLs, activated CTLs and DCs.

\subsection{Preliminaries, assumptions and notation}
\label{sec:preliminariesmeso}
The state of the system at time $t$ is described by the $N$-particle distribution function~\cite{cercignani2013mathematical,villani2002review} 
$$
f^N\left(\mathbf{x}_1, \ldots, \mathbf{x}_N, t,  \mathbf{v}_1, \ldots, \mathbf{v}_N, \mathbf{\tau}_1, \ldots, \mathbf{\tau}_N\right).
$$
In the case where cell dynamics at the microscopic scale obey the rules presented in Section~\ref{Sec:micromodel}, the evolution of $f^N$ is governed by the following transport equation~\cite{kennard1938kinetic}
\begin{equation}
    \partial_t f^N+\sum_{i=1}^{N}\Bigl(\partial_{\tau_i}f^N+c \, \mathbf{v}_i\cdot\nabla_{\mathbf{x}_i} f^N \Bigr)=-\sum_{i=1}^{N}\beta \, f^N \label{eq: trajectories}
\end{equation}
posed on $\Omega^{N} \times \mathbb{R}^*_+ \times {\rm V}^{N} \times \mathbb{R}^{* N}_+$, with 
$$
\Omega^{N} := \{(\mathbf{x}_1, ..., \mathbf{x}_{N}) \in \mathbb{R}^{n\times {N}}:\ |\mathbf{x}_i-\mathbf{x}_j|\geq\varrho \ \forall i,j \} \ .
$$
We consider the transport equation~\eqref{eq: trajectories} subject to smooth, compactly supported initial conditions at $t=0$, boundary conditions corresponding to elastic {interactions} on $\partial\Omega^{N}$, and suitable Dirichlet boundary conditions at $\tau_i=0$ linked to the running probability $\psi$ for $i=1,\ldots,N$. In the mathematical framework given by~\eqref{eq: trajectories}, the probability of finding at position $\mathbf{x}_1$ and at time $t$ the cell labelled by the index $1$ that is moving in direction $\mathbf{v}_1$ for a period of time $\tau_1$ is related to the one-particle marginal
\begin{align*}
{f}(\mathbf{x}_1,t,\mathbf{v}_1,\tau_1)&= \frac{1}{|{\rm V}|^{{ N-1}}}\int_{[0,t]^{{ N-1}}}\int_{\Omega_{N-1}(\mathbf{x}_1)}\int_{{\rm V}^{{ N-1}}}f^{{ N}}(\mathbf{x}_1, \ldots, \mathbf{x}_N, t,  \mathbf{v}_1, \ldots, \mathbf{v}_N, \mathbf{\tau}_1, \ldots, \mathbf{\tau}_N)\nonumber\\  &\ \hspace{6cm} \times\diff\mathbf{v}_{2} \diff\mathbf{x}_{2} \diff\tau_{2} \ldots \diff \mathbf{v}_{N} \diff \mathbf{x}_{N} \diff \tau_{N}\ .
\end{align*}
Here, $|{\rm V}|$ denotes the surface area of the unit sphere ${\rm V}$ and $\Omega_{N-1}(\mathbf{x}_1)\coloneqq \{(\mathbf{x}_{2}, \ldots,\mathbf{x}_{N})  \in \mathbb{R}^{n \times N-1} : (\mathbf{x}_1,\mathbf{x}_{2},\ldots,\mathbf{x}_{N})\in\Omega^{N} \}$.

A comprehensive description of cell dynamics would in principle require considering possible interactions between all cells. However, as mentioned earlier, building on previous work on the mathematical modelling of the interaction dynamics between DCs and CTLs~\cite{macfarlane2019stochastic,macfarlane2018modelling}, we consider only the effect of binary cell-cell interactions, thus neglecting interactions that involve more than two cells. Therefore, as per the scaling and assumptions introduced in Section~\ref{sec: scaling}, which are similar to those typically considered in low-density regimes~\cite{cercignani2013mathematical,othmer1988models,villani2002review}, we truncate the hierarchy of equations corresponding to~\eqref{eq: trajectories} at the second order by integrating out cells $3, \ldots ,N$ from the $N$-particle distribution function $f^N\left(\mathbf{x}_1, \ldots, \mathbf{x}_N, t,  \mathbf{v}_1, \ldots, \mathbf{v}_N, \mathbf{\tau}_1, \ldots, \mathbf{\tau}_N\right)$.

\paragraph{{Two-particle distribution functions}} Let $f_{hk}(\mathbf{x}_h,\mathbf{x}_k,t,\mathbf{v}_h,\mathbf{v}_k,\tau_h,{\tau_k})$ with $h, k \in \{A,D,I\}$ and $k \neq h$ denote the two-particle distribution function associated with: 
\begin{itemize}
\item[-] a cell of population $h$ in the generic phase-space state $(\mathbf{x}_h,\mathbf{v}_h) \in \mathbb{R}^n \times {\rm V}$, with generic run time $\tau_h \in [0,t]$ and stopping rate $ \beta_h(\mathbf{x}_h,\tau_h)$ defined via~\eqref{eq: beta};
\item[-] a cell of population $k$ in the generic phase-space state $(\mathbf{x}_k,\mathbf{v}_k) \in \mathbb{R}^n \times {\rm V}$, with generic run time $\tau_k  \in [0,t]$ and stopping rate $ \beta_k(\mathbf{x}_k,\tau_k)$ defined via~\eqref{eq: beta}.
\end{itemize}
Truncating the hierarchy of equations corresponding to~\eqref{eq: trajectories} at the second order, we obtain the following transport equation for $f_{hk}(\mathbf{x}_h,\mathbf{x}_k,t,\mathbf{v}_h,\mathbf{v}_k,\tau_h,{\tau_k})$
\begin{equation}
\begin{aligned}
    (\partial_t+\partial_{\tau_h}+\partial_{\tau_k} + c \, \mathbf{v}_h\cdot\nabla_{\mathbf{x}_h}+c \, \mathbf{v}_k\cdot\nabla_{\mathbf{x}_k})f_{hk}=-(\beta_h+\beta_k)f_{hk}\label{eq: transpfhkgen}
\end{aligned}
\end{equation}
posed on $\Omega^2 \times \mathbb{R}_+ \times {\rm V}^{2} \times \mathbb{R}^{* 2}_+$, with
\begin{equation}
\label{def:Omegaij}
\Omega^2 := \{(\mathbf{x}_h, \mathbf{x}_{k}) \in \mathbb{R}^{n\times {2}}:\ |\mathbf{x}_h-\mathbf{x}_k|\geq\varrho \ \ \forall h,k \} \ .
\end{equation}
This equation is subject to a smooth, compactly supported initial condition at $t=0$, specular reflective boundary conditions corresponding to elastic interactions on $\partial \Omega^2$, and with boundary conditions  at $\tau_k=0$ and $\tau_h=0$ given by
\begin{equation}
\begin{aligned}
    f_{hk}(\mathbf{x}_h,\mathbf{x}_k,t,\mathbf{v}_h,\mathbf{v}_k,\tau_h,\tau_k=0)&=\mathcal{T}\int_0^t\beta_kf_{hk}(\mathbf{x}_h,\mathbf{x}_k,t,\mathbf{v}_h,\mathbf{v}_k,\tau_h,{\tau_k})\diff \tau_k\ ,\\ f_{hk}(\mathbf{x}_h,\mathbf{x}_k,t,\mathbf{v}_h,\mathbf{v}_k,\tau_h=0,\tau_k)&=\mathcal{T}\int_0^t\beta_hf_{hk}(\mathbf{x}_h,\mathbf{x}_k,t,\mathbf{v}_h,\mathbf{v}_k,\tau_h,{\tau_k})\diff \tau_h\ .\label{boundary conditions for tau}
\end{aligned}
\end{equation}

\paragraph{One-particle distribution functions} Given the two-particle distribution function 
\begin{equation}
\tilde{\tilde{f}}_{hk}(\mathbf{x}_h,\mathbf{x}_k,t,\mathbf{v}_h,\mathbf{v}_k):=\int_0^t\int_0^tf_{hk}\diff\tau_h \diff\tau_k \ , \label{eq: kinetic13ij}
\end{equation}
the one-particle distribution function of population $h$ is given by
\begin{equation}
p_h(\mathbf{x}_h,t,\mathbf{v}_h):=\frac{1}{|{\rm V}|}\int_{\Omega_k(\mathbf{x}_h)}\int_{\rm V}\tilde{\tilde{f}}_{hk}\diff\mathbf{v}_k\diff\mathbf{x}_k \ ,  \label{eq: p densityhk}
    \end{equation}
with the set $\Omega_k(\mathbf{x}_h)$ defined via~\eqref{Omegaxi}. The function $p_h(\mathbf{x}_h,t,\mathbf{v}_h)$ describes the density of cells of population $h$ which at position $\mathbf{x}_h$ and time $t$ move with velocity $\mathbf{v}_h$ (\emph{i.e.} the quantity $p_h(\mathbf{x}_h,t,\mathbf{v}_h) \diff\mathbf{v}_h\diff\mathbf{x}_h$ is the number of cells of population $h$ in the volume element $ \diff\mathbf{v}_h\diff\mathbf{x}_h$ centred at the point $(\mathbf{x}_h, \mathbf{v}_h)$ of the phase space). Moreover, we will consider the weighted two-particle distribution function given by
\begin{equation}
\tilde{\tilde{f}}^{\beta_z}_{hk}(\mathbf{x}_h,\mathbf{x}_k,t,\mathbf{v}_h,\mathbf{v}_k):=\int_0^t\int_0^t \beta_z \ f_{hk}\diff\tau_h \diff\tau_k \ , \quad z \in \{h, k \} \ ,
    \label{eq: kinetic13ijbeta}
\end{equation}
and the weighted one-particle distribution function given by
\begin{equation}
p^{\beta_h}_{h}(\mathbf{x}_h,t,\mathbf{v}_h) :=\frac{1}{|{\rm V}|}\int_{\Omega_k(\mathbf{x}_h)} \ \int_{\rm V} \ \tilde{\tilde{f}}^{\beta_h}_{hk} \diff\mathbf{v}_k\diff\mathbf{x}_k \ . \label{eq: new densityhkbeta}
\end{equation}
%with the stopping rates $\beta_h(\mathbf{x}_h,\tau_h)$ and $\beta_k(\mathbf{x}_k,\tau_k)$ defined via~\eqref{eq: beta}. 
Here $|{\rm V}|$ denotes the surface area of the unit sphere ${\rm V}$.

\subsection{Derivation of a system of transport equations}
\label{Sec:transpeqI}
\paragraph{Transport equations for two-particle distribution functions} The dynamics of the two-particle distribution functions $f_{ID}$ and $f_{AD}$  are governed by the following specific forms of transport equation~\eqref{eq: transpfhkgen}
\begin{align}
    (\partial_t+\partial_{\tau_I}+\partial_{\tau_D} + c \, \mathbf{v}_I\cdot\nabla_{\mathbf{x}_I}+c \, \mathbf{v}_D\cdot\nabla_{\mathbf{x}_D})f_{ID}&=-(\beta_I+\beta_D)f_{ID}\ ,\label{eq: levy initial}\ \\[5pt]
    (\partial_t+\partial_{\tau_A}+\partial_{\tau_D}+c \, \mathbf{v}_A\cdot\nabla_{\mathbf{x}_A}+c \, \mathbf{v}_D\cdot\nabla_{\mathbf{x}_D})f_{AD}&= -(\beta_A+\beta_D)f_{AD}\ , \label{eq: first phase of 2}
    %\\[5pt]
    %(\partial_t+\partial_{\tau_D}+\partial_{\tau_I}+c \, \mathbf{v}_D\cdot\nabla_{\mathbf{x}_D}+c \, \mathbf{v}_I\cdot\nabla_{\mathbf{x}_I})f_{DI}&=-(\beta_I+\beta_D)f_{DI} \ ,  \label{eq: run phase}
\end{align}
which are posed on $\Omega^2 \times \mathbb{R}^*_+ \times {\rm V}^{2} \times \mathbb{R}^{* 2}_+$. The boundary conditions at $\tau_A=0,\ \tau_D=0$ and $\tau_I=0$ are analogous to~\eqref{boundary conditions for tau}. Starting from transport equations~\eqref{eq: levy initial}-\eqref{eq: first phase of 2} and using the method employed in~\cite{estrada2020interacting}, it is possible to show (see Appendix~\ref{sec: appendix A}) that the two-particle distribution functions $\tilde{\tilde{f}}_{ID}$ and $\tilde{\tilde{f}}_{AD}$  given by~\eqref{eq: kinetic13ij} satisfy the following transport equations
\begin{align}
    (\partial_t+c \ \mathbf{v}_I\cdot\nabla_{\mathbf{x}_I} +c \ \mathbf{v}_D\cdot\nabla_{\mathbf{x}_D})\tilde{\tilde{f}}_{ID} =&-(\mathds{1}-\mathcal{T}_I)[\tilde{\tilde{f}}^{\beta_I}_{ID}] -(\mathds{1}-\mathcal{T}_D)[\tilde{\tilde{f}}^{\beta_D}_{ID}]\ , \label{eq: 13 equation}\\[5pt]
    (\partial_t+c \ \mathbf{v}_A\cdot\nabla_{\mathbf{x}_A} +c \ \mathbf{v}_D\cdot\nabla_{\mathbf{x}_D})\tilde{\tilde{f}}_{AD} =&-(\mathds{1}-\mathcal{T}_A)[\tilde{\tilde{f}}^{\beta_A}_{AD}] -(\mathds{1}-\mathcal{T}_D)[\tilde{\tilde{f}}^{\beta_D}_{AD}]\ , \label{eq: 13 equationg}
\end{align}
%and
%\begin{align}
 %   (\partial_t+c \, \mathbf{v}_D\cdot\nabla_{\mathbf{x}_D} +c \, \mathbf{v}_I\cdot\nabla_{\mathbf{x}_I})\tilde{\tilde{f}}_{DI} =&-(\mathds{1}-\mathcal{T}_D)[\tilde{\tilde{f}}^{\beta_D}_{DI}] -(\mathds{1}-\mathcal{T}_I)[\tilde{\tilde{f}}^{\beta_I}_{DI}]  \ ,
  %  \label{eq: 13 equationgCA}
%\end{align}
posed on $\Omega^2 \times \mathbb{R}^*_+ \times {\rm V}^{2}$. Here, $\mathcal{T}_I$, $\mathcal{T}_D$ and $\mathcal{T}_A$ are the turning operators defined via~\eqref{eq: turn angle operator}, and $\tilde{\tilde{f}}^{\beta_h}_{hk}$ and $\tilde{\tilde{f}}^{\beta_k}_{hk}$ are the weighted two-particle distribution functions given by~\eqref{eq: kinetic13ijbeta}. 

\begin{rem}
Notice that the equation describing  { the evolution of the one-particle distribution function $p_D$} will be derived from { the transport equation~\eqref{eq: 13 equation} for the two-particle distribution function $\tilde{\tilde{f}}_{ID}$} by integrating  the variables corresponding to the  $I$  cell, and using the interaction rules  described in Assumption~\ref{Ass1}.
\end{rem}

\paragraph{Transport equation for $p_h$} Starting from transport equation~\eqref{eq: transpfhkgen} and building upon the method presented in~\cite{estrada2020interacting}, it is possible to show (see Appendix~\ref{sec:appendixB}) that the one-particle distribution function $p_h(\mathbf{x}_h,t,\mathbf{v}_h)$ given by~\eqref{eq: p densityhk}  satisfies the following transport equation 
\begin{equation}
    \partial_tp_h+c \, \mathbf{v}_h\cdot\nabla_{\mathbf{x}_h}p_h=-(\mathds{1}-\mathcal{T}_h)[p^{\beta_h}_{h}] + \mathcal{Q}_{hk}, \quad \mathbf{x}_h \in \mathbb{R}^n, t \in \mathbb{R}_+, \mathbf{v}_h \in {\rm V} \ .
     \label{eq:transpeqph}
\end{equation}
Here, the turning operator $\mathcal{T}_h$ is defined via~\eqref{eq: turn angle operator}, the weighted one-particle distribution function $p^{\beta_h}_{h}(\mathbf{x}_h,t,\mathbf{v}_h)$ is given by~\eqref{eq: new densityhkbeta} and
\begin{equation}
\label{def:Qhk}
\mathcal{Q}_{hk}(\mathbf{x}_h,t,\mathbf{v}_h) :=\dfrac{c}{|{\rm V}|}\int_{\partial {\rm B}_\varrho(\mathbf{x}_h)}\int_{\rm V}\nu\cdot(\mathbf{v}_h-\mathbf{v}_k)\tilde{\tilde{f}}_{hk} \diff\mathbf{v}_k\diff\sigma \ .
\end{equation}
{ In~\eqref{def:Qhk}, $\nu$ is the unit normal defined in~\eqref{eq: new velocity} and $\diff \sigma$ denotes the surface element.} 

{ The first term on the right-hand side of transport equation~\eqref{eq:transpeqph} represents the rate of change of the one-particle distribution function due to cell movement, while the term $\mathcal{Q}_{hk}$ is the rate of change due to interactions between cells. The specific forms of these terms depend, respectively, on the way in which cells move and the interactions they undergo, as discussed in the remainder of this section.}

{ \paragraph{Expressions for $p^{\beta_h}_{h}$} The specific form of the first term on the right-hand side of transport equation~\eqref{eq:transpeqph} depends on the expression for $p^{\beta_h}_{h}$ which, in turn, will depend on the definition of the stopping rate $\beta_h$.} 

When cells move in a local search pattern (i.e. for $h=A$ and $h=D$), the stopping rate $\beta_h$ is defined via~\eqref{eq: beta} and~\eqref{eq: brownian}. In this case, inserting the definition of $\beta_h$ into~\eqref{eq: new densityhkbeta} yields 
\begin{equation}
\label{eq:phbetabro}
p_h^{\beta_h}(\mathbf{x}_h,t,\mathbf{v}_h)=b \, p_h(\mathbf{x}_h,t,\mathbf{v}_h) \ .
\end{equation}
On the other hand, when cells move in a non-local search pattern (i.e. for $h=I$), the stopping rate $\beta_h$ is defined via~\eqref{eq: beta} and~\eqref{eq: levy}. In this case, it is possible to show (see Appendix~\ref{sec:appendixC}) that 
\begin{equation}
\label{eq:phbetalev}
 p^{\beta_h}_h(\mathbf{x}_h,t,\mathbf{v}_h)=\mathcal{B}[p_h](\mathbf{x}_h,t,\mathbf{v}_h) \ ,
\end{equation}
where $\mathcal{B}$ is a convolution operator such that
\begin{equation}\label{def:operatorB}
\mathcal{B}[p_h](\mathbf{x}_h,t, \mathbf{v}_h) = \int_0^t B(\mathbf{x}_h,t-s)p(\mathbf{x}_h-(c\mathbf{v}_h+b)(t-s),s,\mathbf{v}_h)\diff s \ ,
\end{equation}
with $B$ being defined through its Laplace transform in time $\hat{B}$ as
\begin{equation}
    \hat{B}(\mathbf{x}_h,\lambda+b+c\mathbf{v}_h\cdot\nabla_{\mathbf{x}_h})=\frac{\hat{\varphi}_h(\mathbf{x}_h,\lambda+b+c\mathbf{v}_h\cdot\nabla_{\mathbf{x}_h})}{\hat{\psi}_h(\mathbf{x}_h,\lambda+b+c\mathbf{v}_h\cdot\nabla_{\mathbf{x}_h})}\ .\label{eq: turn angle operator B}
\end{equation}
Here, $\lambda$ is the Laplace variable, $\hat{\varphi}_h$ and $\hat{\psi}_h$ are the Laplace transforms in $\tau_h$ of the functions $\varphi_h$ and $\psi_h$ defined via~\eqref{eq: levy}, and the parameter $b$ is defined via~\eqref{eq: brownian}. 

\paragraph{Expressions for $\mathcal{Q}_{hk}$} Following~\cite{estrada2020interacting,franz2016hard}, we first note that when a  cell in the phase-space state $({\bf x}_h, {\bf v}_h)$ interacts with a cell in the phase-space state $({\bf x}_k, {\bf v}_k)$ we have $|{\bf x}_h - {\bf x}_k| = \varrho$. Hence, the normal vector at the point of physical contact between the interacting cells, $\nu \in {\rm V}$, defined via~\eqref{eq: new velocity} can be written as $\nu=(\mathbf{x}_h-\mathbf{x}_k)/\varrho$, that is, $\mathbf{x}_k=\mathbf{x}_h-\nu\varrho$. As a result, using the fact that ${\rm B}_\varrho = \varrho {\rm V}$ along with the change of variable $\nu \mapsto -\nu$, we rewrite~\eqref{def:Qhk} as 
\begin{align}
\mathcal{Q}_{hk}(\mathbf{x}_h,t,\mathbf{v}_h) := & \frac{c}{|{{\rm V}}|}\int_{\partial {\rm B}_\varrho(\mathbf{x}_h)}\int_{\rm V}\nu\cdot(\mathbf{v}_h-\mathbf{v}_k)\tilde{\tilde{f}}_{hk} \diff\mathbf{v}_k\diff\sigma \nonumber \\ = & -\frac{c}{|{{\rm V}}|}\varrho^{n-1}\int_{{\rm V}}\int_{\rm V}\nu\cdot(\mathbf{v}_h-\mathbf{v}_k)\tilde{\tilde{f}}_{hk}(\mathbf{x}_h,\mathbf{x}_h+\nu\varrho,t,\mathbf{v}_h,\mathbf{v}_k)\diff\mathbf{v}_k\diff\nu\ .\label{eq: interaction term for N}
\end{align}
Following~\cite{estrada2020interacting}, we also note that ${\rm V} \equiv {\rm V}^+_{hk}\cup {\rm V}^-_{hk}$ with
\begin{equation}
\begin{aligned}
\label{def:Vhk}
{\rm V}^+_{hk} &\coloneqq \{\nu \in {\rm V} : \nu\cdot(\mathbf{v}_h-\mathbf{v}_k)>0 \}\ , \\
{\rm V}^-_{hk} &\coloneqq \{\nu \in {\rm V} : \nu\cdot(\mathbf{v}_h-\mathbf{v}_k)<0 \} \equiv \{-\nu \in {\rm V} : \nu\cdot(\mathbf{v}_h-\mathbf{v}_k)>0 \}\ .
\end{aligned}
\end{equation}
Therefore, a cell moving in direction ${\bf v}_h$ and a  cell moving in direction ${\bf v}_k$ will move toward each other if $\nu \in {\rm V}^+$ and away from each other if $\nu \in {\rm V}^-$. 

{Under Assumptions~\ref{Ass1}-\ref{Ass2}, denoting the post-{interaction} directions corresponding to $\mathbf{v}_h$ and $\mathbf{v}_k$ by $\mathbf{v}'_h$ and $\mathbf{v}'_k$, which are defined via~\eqref{eq: new velocity}, we consider two different types of interactions between cells:
\begin{itemize}
\item[-]  { conservative interactions} { (\emph{cf.} Definition~\ref{Ass0})}, between a  cell of population $h$ and a  cell of population $k$, whereby both cells remain in their original populations upon interaction and acquire the post-{interaction} velocities;
\item[-]  population-switching interactions { (\emph{cf.} Definition~\ref{Ass0b})}, between a cell in population $h$ and a  cell in population $k$, whereby the $h$ cell switches from its original population to a different one upon interaction.
\end{itemize}}

{ From~\eqref{eq: interaction term for N}, we define the rate of change of the one-particle distribution function $p_h(\mathbf{x}_h,t,\mathbf{v}_h)$ due to conservative interactions as 
 \begin{align}
     \mathcal{K}_{hk}(\mathbf{x}_h,t,\mathbf{v}_h) & :=-\frac{c}{|{{\rm V}}|}\varrho^{n-1} \, N_k(t) \, \Big[\int_{{\rm V}^+_{hk}}\int_{\rm V}\nu\cdot(\mathbf{v}_h-\mathbf{v}_k)\tilde{\tilde{f}}_{hk}(\mathbf{x}_h,\mathbf{x}_h+\nu\varrho,t,\mathbf{v}_h,\mathbf{v}_k)\diff\mathbf{v}_k\diff\nu\nonumber\\& \qquad +\int_{{\rm V}^-_{hk}}\int_{\rm V}\nu\cdot(\mathbf{v}_h-\mathbf{v}_k)\tilde{\tilde{f}}_{hk}(\mathbf{x}_h,\mathbf{x}_h+\nu\varrho,t,\mathbf{v}_h,\mathbf{v}_k)\diff\mathbf{v}_k\diff\nu\Big]\nonumber\\
     &=\frac{c}{|{{\rm V}}|}\varrho^{n-1} \, N_k(t) \, \int_{{\rm V}_{hk}^+}\int_{\rm V}\nu\cdot(\mathbf{v}_h-\mathbf{v}_k)\Bigl[\tilde{\tilde{f}}_{hk}(\mathbf{x}_h,\mathbf{x}_h-\nu\varrho,t,\mathbf{v}'_h,\mathbf{v}'_k)\nonumber\\ &\qquad -\tilde{\tilde{f}}_{hk}(\mathbf{x}_h,\mathbf{x}_h+\nu\varrho,t,\mathbf{v}_h,\mathbf{v}_k)\Bigr]\diff\mathbf{v}_k\diff\nu\ ,\label{def:Jhk}
 \end{align} 
with $N_k(t)$ being the number of cells in population $k$ at time $t$. The second equality in~\eqref{def:Jhk} is obtained by using the normal vector $-\nu$ and the post-{interaction} directions $\mathbf{v}_h'$ and $\mathbf{v}_k'$ in $\tilde{\tilde{f}}_{hk}$ over the set ${\rm V}_{hk}^-$. Notice that the following property holds
\begin{equation}
\label{eq:propKhk}
\int_{\rm V} \mathcal{K}_{hk}(\cdot,\cdot,\mathbf{v}_h) \diff\mathbf{v}_h = 0 \ , 
\end{equation}
which ensures that the density of cells in population $h$ will be preserved in the course of such interactions.
}

%\end{equation}

{Moreover, based on~\eqref{eq: interaction term for N} and~\eqref{def:Jhk}, we define the rate of change of the one-particle distribution function $p_h(\mathbf{x}_h,t,\mathbf{v}_h)$ due to population-switching interactions  leading the cell to leave population $h$ as
\begin{equation}
  \mathcal{J}_{hk}(\mathbf{x}_h,t,\mathbf{v}_h) := - \dfrac{c}{|{\rm V}|} \, \varrho^{n-1} \, N_k(t) \, \int_{{\rm V}^+_{hk}}\int_{\rm V}\nu\cdot(\mathbf{v}_h-\mathbf{v}_k) \, \tilde{\tilde{f}}_{hk}(\mathbf{x}_h,\mathbf{x}_h+\nu\varrho, t,\mathbf{v}_h,\mathbf{v}_k)\diff\mathbf{v}_k\diff\nu \ . \label{def:Jhto}
\end{equation}
}
{Analogously, we define the rate of change of  $p_h(\mathbf{x}_h,t,\mathbf{v}_h)$ due to population-switching interactions leading a cell to leave a generic population $l \neq h$ and enter population $h$ as
\begin{align}
\label{def:opm}
\mathcal{J}^h_{lk}(\mathbf{x}_h,t,\mathbf{v}_h) :=  &\dfrac{c}{|{\rm V}|} \, \varrho^{n-1}  \, N_k(t) \, \int_{ \Omega_l(\mathbf{x}_k)}\int_{\rm V} \delta(\mathbf{x}_l - \mathbf{x}_h) \, \delta(\mathbf{v}_l - \mathbf{v}_h)\nonumber \\
&\quad\times \int_{{\rm V}^+_{lk}}\int_{\rm V}\nu\cdot(\mathbf{v}_l-\mathbf{v}_k) \tilde{\tilde{f}}_{lk}(\mathbf{x}_l,\mathbf{x}_l+\nu\varrho, t, \mathbf{v}_l,\mathbf{v}_k) \diff\mathbf{v}_k\diff\nu \diff\mathbf{v}_l\diff\mathbf{x}_l \ ,
\end{align}
with $\delta({\mathbf{z}} - {\mathbf{z}}^*)$ being the Dirac delta distribution centred at ${\mathbf{z}}^*$. 
{Definition~\eqref{def:opm} ensures that the density of cells that leave population $l$ due to such interactions will appear in population $h$. In fact, we have}
\begin{equation*}
    \mathcal{J}_{lk}^h(\mathbf{x}_h,t,\mathbf{v}_h)=-\int_{\Omega_l(\mathbf{x}_k)}\int_{\rm V}\delta(\mathbf{x}_l-\mathbf{x}_h)\delta(\mathbf{v}_l-\mathbf{v}_h)\mathcal{J}_{lk}(\mathbf{x}_l,t,\mathbf{v}_l)\diff\mathbf{v}_l\diff\mathbf{x}_l\ .
\end{equation*}

In summary, the term $\mathcal{Q}_{hk}$ in transport equation~\eqref{eq:transpeqph} is defined in terms of~\eqref{def:Jhk}-\eqref{def:opm} in different possible ways depending on the cell-cell interactions that are considered.}

{ Under Assumptions~\ref{Ass1}-\ref{Ass2} and definitions~\eqref{def:Jhk}, \eqref{def:Jhto} and~\eqref{def:opm}, the rates of change of the one-particle distribution functions $p_I(\mathbf{x}_I,t,\mathbf{v}_I)$, $p_A(\mathbf{x}_A,t,\mathbf{v}_A)$ and $p_D(\mathbf{x}_D,t,\mathbf{v}_D)$ due to cell-cell interactions will be, respectively,
\begin{align}
\label{eq:colliratesID}
\mathcal{Q}_{ID}(\mathbf{x}_I,t,\mathbf{v}_I)& = (1-\zeta)\mathcal{K}_{ID}(\mathbf{x}_I,t,\mathbf{v}_I) {+} \zeta \mathcal{J}_{ID}(\mathbf{x}_I,t,\mathbf{v}_I) \ ,\\[5pt]
\label{eq:colliratesAD}
\mathcal{Q}_{AD}(\mathbf{x}_A,t,\mathbf{v}_A) &= \mathcal{K}_{AD}(\mathbf{x}_A,t,\mathbf{v}_A) + \zeta \, \mathcal{J}^A_{ID}(\mathbf{x}_A,t,\mathbf{v}_A) \ , \\[5pt]
\label{eq:colliratesDI}
\mathcal{Q}_{DI}(\mathbf{x}_D,t,\mathbf{v}_D) &= \mathcal{K}_{DI}(\mathbf{x}_D,t,\mathbf{v}_D) \ .
\end{align}
}
%{\red Here, we have introduced a parameter $r$ which describes the rate of occurrence of the switching interactions represented by $\mathcal{J}_{ID}$ and $\mathcal{J}_{ID}^A$.}\\
%{\magenta Here, we have introduced a parameter $r$ which captures possible differences in the rate at which interactions that may result in CTL activation occur compared to other cell-cell interactions.}\\

{ Substituting~\eqref{eq:phbetabro}, \eqref{eq:phbetalev} and~\eqref{eq:colliratesID}-\eqref{eq:colliratesDI} into transport equation~\eqref{eq:transpeqph}, we obtain} the following transport equations for  $p_I(\mathbf{x}_I,t,\mathbf{v}_I)$, $p_A(\mathbf{x}_A,t,\mathbf{v}_A)$ and $p_D(\mathbf{x}_D,t,\mathbf{v}_D)$:
\begin{align}
    \partial_tp_I+c \, \mathbf{v}_I\cdot\nabla_{\mathbf{x}_I}p_I =&\underbrace{-(\mathds{1}-\mathcal{T}_I)\mathcal{B}[p_I]}_{\mbox{\scriptsize{cell motion}}} + \underbrace{(1-\zeta)\mathcal{K}_{ID}}_{\mbox{\scriptsize{interactions}}} \nonumber \\ & \qquad \qquad \qquad \underbrace{{+} \, \zeta \mathcal{J}_{ID},}_{\substack{\mbox{\scriptsize{outflow due}}\\\mbox{\scriptsize{to activation}}}} \quad \mathbf{x}_I \in \mathbb{R}^n, t \in \mathbb{R}^*_+,\mathbf{v}_I \in {\rm V} \ , \label{eq: levy kinetic equation}
\end{align}
\begin{align}
    \partial_t{p}_A+c \, \mathbf{v}_A\cdot\nabla_{\mathbf{x}_A}{p}_A=& \underbrace{-b \, (\mathds{1}-\mathcal{T}_A)[p_{A}]}_{\mbox{\scriptsize{cell motion}}}  + \underbrace{\mathcal{K}_{AD}}_{\mbox{\scriptsize{interactions}}} \nonumber \\ & \quad \underbrace{+ \zeta \, \mathcal{J}^A_{ID},}_{\substack{\mbox{\scriptsize{inflow due}}\\\mbox{\scriptsize{to activation}}}} \quad \hspace{1.8cm} \mathbf{x}_A \in \mathbb{R}^n, t \in \mathbb{R}^*_+,\mathbf{v}_A \in {\rm V}\ , \label{eq: kinetic equation for brownian}
    \end{align}
    \begin{align}
    \partial_t {p}_D +c\mathbf{v}_D\cdot\nabla_{\mathbf{x}_D}{p}_D = &\underbrace{-b \, (\mathds{1}-\mathcal{T}_D)[p_{D}]}_{\mbox{\scriptsize{cell motion}}} \nonumber \\ & \quad \underbrace{+\mathcal{K}_{DI},}_{\mbox{\scriptsize{interactions}}} \quad \hspace{1.8cm} \mathbf{x}_D \in \mathbb{R}^n, t \in \mathbb{R}^*_+,\mathbf{v}_D \in {\rm V}  \ . \label{eq: kinetic cancer without proliferation}
\end{align}
The terms on the right-hand sides of~\eqref{eq: levy kinetic equation}-\eqref{eq: kinetic cancer without proliferation} represent the rate of change of the one-particle distributions due to the biophysical phenomena specified below each term.

\section{Macroscopic scale model}
\label{Sec:macromodel}
In this section, we derive a macroscopic system of equations corresponding to the mescoscopic scale model given by transport equations \eqref{eq: levy kinetic equation}-\eqref{eq: kinetic cancer without proliferation}. Such a model consists of a coupled system of balance equations for the macroscopic densities of inactive CTLs, activated CTLs and DCs.
 
\subsection{Preliminaries, assumptions and notation}
\label{sec: scaling}
\paragraph{Scaling} We assume the mean run time $\bar \tau$ to be small compared to the characteristic temporal scale for the dynamics of the macroscopic cell densities, which is represented by the parameter $T \in \mathbb{R}^*_+$, i.e. we make the assumption 
$$
\dfrac{\bar \tau}{T} =: \e \ll 1 \ .
$$
Moreover, we let $X \in \mathbb{R}^*_+$ represent the characteristic spatial scale for the dynamics of the macroscopic cell densities and introduce the rescaled quantities
$$
\hat{t} = \dfrac{t}{T}, \quad \hat{{\bf x}} = \dfrac{{\bf x}}{X}, \quad \hat{\tau} = \dfrac{\bar \tau}{T}, \quad \hat{c} = c \, \dfrac{T}{X} \ .
$$

As similarly done in~\cite{alt1980biased,estrada2020interacting}, in order to obtain a mathematical model for the dynamics of the cells at the macroscopic scale, we consider the scaling
\begin{equation}
\label{def:scaling}
 (\mathbf{x},t,c,\tau)\mapsto(\hat{\mathbf{x}}/\varepsilon,\hat{t}/\varepsilon,\hat{c}/\varepsilon^\gamma,\hat{\tau}/\varepsilon^\mu) \ ,
\end{equation}
with
\begin{equation}
\label{ass:gammamu}
\gamma, \mu \in \mathbb{R}^*_+ \ , \quad \gamma < 1 \quad \text{and} \quad \mu > 1 - \gamma \ .
\end{equation}
Throughout the rest of the paper, we will drop the carets from~\eqref{def:scaling} and we will study two-dimensional cell dynamics (i.e. we assume $n=2$). 

Furthermore, noting that the diameter of the cells is small compared to the characteristic spatial scale for the dynamics of the macroscopic cell densities, and considering a biological scenario where the number of cells in the system is large and activation of CTL occurs with a small probability $\zeta$, we assume
\begin{equation}
\label{def:scalingrhoN}
\varrho = \e^{\xi} \ , \quad N_I(t) \equiv \varepsilon^{-\vartheta}, \quad  N_D = \varepsilon^{-\vartheta}\ , \quad \zeta = \e^\kappa \ , \quad \xi, \vartheta, \kappa \in \mathbb{R}^*_+ \ .
\end{equation}
In particular, we will be focussing on a biological scenario corresponding to the following assumptions
\begin{equation}
\label{ass:scal1}
\gamma := \dfrac{1}{2} \ , \quad \xi-\vartheta := 1 - \dfrac{\gamma}{\alpha -1} \quad \text{and} \quad \kappa = - (\xi-\vartheta) + \dfrac{3}{2} > 0 \ .
%\upxi-\vartheta > \max \left(\gamma, \gamma -1 \right) = \dfrac{1}{2} 
\end{equation}
Notice that $\xi-\vartheta<0$ when $\alpha<3/2$. In the case where cells follow a Brownian motion (i.e. for $h=A$ and $h=D$)  we have $\alpha=2$ and, therefore, $\xi-\vartheta=1-\gamma=1/2$.
Under scaling~\eqref{def:scaling} definitions~\eqref{eq: levy} become
\begin{equation}
    \psi^\e(\mathbf{x}_i,\tau_i)=\Bigl(\frac{\varepsilon^\mu\tau_0({\bf x}_i)}{\varepsilon^\mu\tau_0({\bf x}_i)+\tau_i}\Bigr)^\alpha\ , \quad \varphi^\e(\mathbf{x}_i,\tau_i):=\frac{\alpha \, \varepsilon^\mu \, \tau_0(\mathbf{x}_i)^\alpha}{(\varepsilon^\mu \, \tau_0(\mathbf{x}_i)+\tau_i)^{\alpha+1}}, \quad \alpha\in(1,2) \ .
    \label{eq: levyresc} 
\end{equation}
Moreover, under assumption~\eqref{def:scalingrhoN} on $\varrho$ we have
\begin{equation}\label{eq:ftildetilderesc}
 \tilde{\tilde{f}}_{hk}(\mathbf{x}_h,\mathbf{x}_h \pm \nu \rho,t,\mathbf{v}_h,\mathbf{v}_k) \equiv \tilde{\tilde{f}}_{hk}(\mathbf{x}_h,\mathbf{x}_h \pm\varepsilon^{\upxi}\nu,t,\mathbf{v}_h,\mathbf{v}_k) \ .
\end{equation}

\paragraph{``Molecular chaos'' assumption} Considering a biological scenario where cell densities are sufficiently low, we assume the velocities of any two cells which are about to {interact} to be uncorrelated -- \emph{i.e.} we make the so-called ``molecular chaos'' assumption, which holds at low densities and is commonly used in kinetic theory~\cite{cercignani2013mathematical,othmer1988models,villani2002review}. Under this assumption, the two-particle distribution function $\tilde{\tilde{f}}_{hk}(\mathbf{x}_h,\mathbf{x}_h \pm\varepsilon^{\upxi}\nu,t,\mathbf{v}_h,\mathbf{v}_k)$ can be approximately expressed as the product of the corresponding one-particle distribution functions, that is, 
\begin{equation}\label{eq:ftildetildefact}
    \tilde{\tilde{f}}_{hk}(\mathbf{x}_h,\mathbf{x}_h \pm\varepsilon^{\upxi}\nu,t,\mathbf{v}_h,\mathbf{v}_k)=p_{h}^\e(\mathbf{x}_h,t,\mathbf{v}_h) \, p_{k}^\e(\mathbf{x}_h,t,\mathbf{v}_k)+\mathcal{O}(\varepsilon^{\upxi})\ .
\end{equation}
{We draw the attention of the reader to the fact that, throughout the rest of the paper, superscript and subscript related to the scaling should not be confused with the index of another cell population.}

%\paragraph{Rescaled interaction terms}  
Under scaling~\eqref{def:scaling} and assumptions~\eqref{def:scalingrhoN}, using \eqref{eq:ftildetilderesc}, \eqref{eq:ftildetildefact} and assuming $n=2$, the interaction terms defined via~\eqref{def:Jhk}, \eqref{def:Jhto} and~\eqref{def:opm} read as

 \begin{align}
     \mathcal{K}_{hk}^\e(\mathbf{x}_h,t,\mathbf{v}_h)& = \frac{1}{|{{\rm V}}|} \e^{\xi -\vartheta - \gamma} \,c \, \int_{{\rm V}_{hk}^+}\int_{\rm V}\nu\cdot(\mathbf{v}_h-\mathbf{v}_k)\Bigl[p_{h}^\e(\mathbf{x}_h,t,{\bf v}'_h) p_{k}^\e(\mathbf{x}_h,t,{\bf v}'_k) \nonumber\\ &\qquad\qquad - p_{h}^\e(\mathbf{x}_h,t,{\bf v}_h) p_{k}^\e(\mathbf{x}_h,t,{\bf v}_k)\Bigr]\diff\mathbf{v}_k\diff\nu\ ,\label{def:Jhkeps}\\[5pt]
  \mathcal{J}_{hk}^\e(\mathbf{x}_h,t,\mathbf{v}_h)&= - \frac{1}{|{{\rm V}}|} \e^{\xi -\vartheta - \gamma} \, c \, \int_{{\rm V}^+_{hk}}\int_{\rm V}\nu\cdot(\mathbf{v}_h-\mathbf{v}_k) \, p_{h}^\e(\mathbf{x}_h,t,{\bf v}_h) p_{k}^\e(\mathbf{x}_h,t,{\bf v}_k)\diff\mathbf{v}_k\diff\nu \label{def:Jhtoeps}
\end{align}
and
\begin{align}
\label{def:opmeps}
%{\blue \dfrac{1}{|{\rm V}|}} \,
{}_{\e}\mathcal{J}^h_{lk }(\mathbf{x}_h,t,\mathbf{v}_h) =  &\dfrac{1}{|{\rm V}|} \, \e^{\xi -\vartheta - \gamma} \, c \, \int_{ \Omega_l(\mathbf{x}_k)}\int_{\rm V} \delta(\mathbf{x}_l - \mathbf{x}_h) \, \delta(\mathbf{v}_l - \mathbf{v}_h)\nonumber \\
&\quad\times \int_{{\rm V}^+_{lk}}\int_{\rm V}\nu\cdot(\mathbf{v}_l-\mathbf{v}_k) p_{l}^\e(\mathbf{x}_l,t,{\bf v}_l) p_{k}^\e(\mathbf{x}_l,t,{\bf v}_k) \diff\mathbf{v}_k\diff\nu \diff\mathbf{v}_l\diff\mathbf{x}_l \ .
\end{align}

\paragraph{Expansion of $p_{h}^\e$ and macroscopic cell quantities}  
Exploiting the results established by Lemma~\ref{lem: eigenfunctions} in the case where $n=2$, we expand the one-particle distribution function $p_{h}^\e$ in terms of its zeroth moment $\rho_{h}^\e$ (i.e. the macroscopic cell density) and its first moment $w_{h}^\e$ (i.e. the local macroscopic direction of cell motion). This is possible because, as one can see from the right-hand side of transport equation~\eqref{eq:transpeqphresc}, the interaction terms are of higher order in $\varepsilon$ (\emph{cf.} the scaling used in \eqref{def:Jhkeps}-\eqref{def:opmeps}) and, therefore, we can write
\begin{equation}\label{eq:phfact}
    p_{h}^\e(\mathbf{x}_h,t,\mathbf{v}_h)=\frac{1}{|{\rm V}|} \Big(\rho_{h}^\e(\mathbf{x}_h,t) + \varepsilon^{\gamma} \,{ 2} \, \mathbf{v}_h\cdot w_{h}^\e(\mathbf{x}_h,t) \Big) + o(\varepsilon^{\gamma}) \ , \quad h \in \{A, D, I \} \ ,
\end{equation}
where
\begin{equation}
\rho_{h}^\e(\mathbf{x}_h,t) := \int_{\rm V} p_{h}^\e(\mathbf{x}_h,t,\mathbf{v}_h)\diff \mathbf{v}_h\ ,\ \ \ w_{h}^\e(\mathbf{x}_h,t) := \frac{1}{\varepsilon^\gamma}\int_{\rm V}\mathbf{v}_h \, p_{h}^\e(\mathbf{x}_h,t,\mathbf{v}_h)\diff\mathbf{v}_h \ .\label{eq: macroscopic quantities}
\end{equation}
 We refer the reader to \cite{othmer2000diffusion,othmer2012experimental} and the seminal work~\cite{alt1980biased}  for a complete derivation in the case of no interactions and to~\cite{estrada2020interacting,franz2016hard} for the case of velocity jump models with interacting particles. The appropriate choice of scaling for the local macroscopic direction of motion is found by first inserting \eqref{eq:phfact} into~\eqref{eq:transpeqphresc} and then integrating over ${\rm V}$ in order to obtain a suitable macroscopic equation (see transport equation~\eqref{eq:transpeqrhoepstemp}).

\subsection{Derivation of a macroscopic scale system}
\paragraph{Transport equation for $p_{h}^\e$} Under scaling~\eqref{def:scaling} and assumptions~\eqref{def:scalingrhoN}, using \eqref{eq:ftildetilderesc}, \eqref{eq:ftildetildefact} and assuming $n=2$, we rewrite transport equation~\eqref{eq:transpeqph} for the one-particle distribution function $p_h(\mathbf{x}_h,t,\mathbf{v}_h)$ as
\begin{equation}
\e  \partial_tp_{h}^\e + \e^{1-\gamma} c \, \mathbf{v}_h\cdot\nabla_{\mathbf{x}_h}p_{h}^\e=-(\mathds{1}-\mathcal{T}_h)[{}_{\e}p^{\beta_h}_{h}]  +  \mathcal{Q}_{hk}^\e \ ,
  \label{eq:transpeqphresc}
\end{equation}
where $\mathcal{Q}_{hk}^\e$ is defined in terms of $\mathcal{K}_{hk}^\e$, $ \mathcal{J}_{hk}^\e$ and ${}_{\e}\mathcal{J}^h_{lk}$ as per~\eqref{eq:colliratesID}-\eqref{eq:colliratesDI}, that is,
\begin{align}
\label{eq:colliratesIDeps}
\mathcal{Q}_{ID}^\e(\mathbf{x}_I,t,\mathbf{v}_I) &= (1-\e^\kappa)\mathcal{K}_{ID}^\e(\mathbf{x}_I,t,\mathbf{v}_I) + \e^\kappa \mathcal{J}_{ID}^\e(\mathbf{x}_I,t,\mathbf{v}_I) \ ,\\[5pt]
\label{eq:colliratesADeps}
\mathcal{Q}_{AD}^\e(\mathbf{x}_A,t,\mathbf{v}_A) &= \mathcal{K}_{AD}^\e(\mathbf{x}_A,t,\mathbf{v}_A) +  \e^\kappa \, {}_{\e}\mathcal{J}^A_{ID}(\mathbf{x}_A,t,\mathbf{v}_A) 
\end{align}
and
\begin{equation}
\label{eq:colliratesDIeps}
\mathcal{Q}_{DI}^\e(\mathbf{x}_D,t,\mathbf{v}_D) = \mathcal{K}_{DI}^\e(\mathbf{x}_D,t,\mathbf{v}_D) \ . 
\end{equation}

We recall that in the case where cells move in a local search pattern (i.e. for $h=A$ and $h=D$), $\beta_h$ is defined via~\eqref{eq: beta} and~\eqref{eq: brownian}, and thus ${}_{\e}p^{\beta_h}_{h}(\mathbf{x}_h,t,\mathbf{v}_h)$ is given as in~\eqref{eq:phbetabro}. On the other hand, in the case where cells move in a non-local search pattern (i.e. for $h=I$), $\beta_h$ is defined via~\eqref{eq: beta} and~\eqref{eq: levy}, and thus ${}_{\e}p^{\beta_h}_{h}(\mathbf{x}_h,t,\mathbf{v}_h)$ is given by~\eqref{eq:phbetalev}
\iffalse
In the case where cells move in a local search pattern (i.e. for $h=A$ and $h=D$), $\beta_h$ is defined via~\eqref{eq: beta} and~\eqref{eq: brownian} and thus we have
\begin{equation}
{}_{\e}p^{\beta_h}_{h}(\mathbf{x}_h,t,\mathbf{v}_h) = b \, p_{h}^\e(\mathbf{x}_h,t,\mathbf{v}_h) \label{eq:Tepsbrownian}
\end{equation}
as per~\eqref{eq:phbetabro}. On the other hand, in the case where cells move in a non-local search pattern (i.e. for $h=I$), $\beta_h$ is defined via~\eqref{eq: beta} and~\eqref{eq: levy} and thus we have
\begin{equation}
{}_{\e}p^{\beta_h}_{h}(\mathbf{x}_h,t,\mathbf{v}_h) = \mathcal{B}^{\e}[p_{h}^\e](\mathbf{x}_h,t,\mathbf{v}_h) \label{eq:Tepslevy}
\end{equation}
as per~\eqref{eq:phbetalev}, 
\fi
with
$$
\mathcal{B}^{\e}[p_{h}^\e](\mathbf{x}_h,t,\mathbf{v}_h) = \int_0^t B^{\e}(\mathbf{x}_h,t-s)p_{h}^\e(\mathbf{x}_h-(c\mathbf{v}_h+b)(t-s),s)\diff s \ .
$$
As before, $B^{\e}$ is defined through its Laplace transform in time $\hat{B}^{\e}$ and, in particular, under assumptions~\eqref{ass:gammamu}, we make the approximation
\[
\hat{B}^\varepsilon(\mathbf{x}_h,\varepsilon\lambda+\varepsilon^\mu b+\varepsilon^{1-\gamma}c\mathbf{v}_h\cdot\nabla_{\mathbf{x}_h})\simeq \hat{B}^\varepsilon(\mathbf{x}_h,\varepsilon^{1-\gamma}c\mathbf{v}_h\cdot\nabla_{\mathbf{x}_h}) \ .\label{eq: quasi-static approximation}
\]
Using the properties of the Laplace transform of a convolution, we write
\[
\int_0^t B^\varepsilon(\mathbf{x}_h,t-s)p_{h}^\e(\mathbf{x}_h-(c\mathbf{v}_h+b)(t-s),s,\mathbf{v}_h)\diff s\simeq \hat{B}^\varepsilon(\mathbf{x}_h,\varepsilon^{1-\gamma}c\mathbf{v}_h\cdot\nabla_{\mathbf{x}_h})p_h^\e(\mathbf{x}_h,t,\mathbf{v}_h)\ ,
\]
with
\begin{equation}
    \hat{B}^\varepsilon(\mathbf{x}_h,\varepsilon^{1-\gamma}c\mathbf{v}_h\cdot\nabla_{\mathbf{x}_h})=\frac{\hat{\varphi}_h^\varepsilon(\mathbf{x}_h,\varepsilon^{1-\gamma}c\mathbf{v}_h\cdot\nabla_{\mathbf{x}_h})}{\hat{\psi}_h^\varepsilon(\mathbf{x}_h,\varepsilon^{1-\gamma}c\mathbf{v}_h\cdot\nabla_{\mathbf{x}_h})} \label{eq: turn angle operator Bepstemp} \ .
\end{equation}
Analogous calculations are fully detailed in Appendix~\ref{sec:appendixC}.
%Here, $\hat{\varphi}_{h}^\e$ and $\hat{\psi}_{h}^\e$ are the Laplace transforms in $\tau_h$ of the functions $\varphi_{h}^\e$ and $\psi_{h}^\e$ defined via~\eqref{eq: levyresc}, and the parameter $b$ is defined via~\eqref{eq: brownian}. 
Substituting the expressions of $\hat{\varphi}_{h}^\e$ and $\hat{\psi}_{h}^\e$ into~\eqref{eq: turn angle operator Bepstemp}, calculations similar to those carried out  in~\cite{estrada2020interacting,estrada2018fractional} allow one to show that
\begin{align}
    \hat{B}^\varepsilon&(\mathbf{x}_h,\varepsilon^{1-\gamma}c\mathbf{v}_h\cdot\nabla_{\mathbf{x}_h})= \frac{\alpha-1}{d_{\e}}-\frac{\varepsilon^{1-\gamma}}{2-\alpha}c\mathbf{v}_h\cdot\nabla_{\mathbf{x}_h}\nonumber\\ &-d_{\e}^{\alpha-2}\varepsilon^{(1-\gamma)(\alpha-1)}(c\mathbf{v}_h\cdot\nabla_{\mathbf{x}_h})^{\alpha-1}(\alpha-1)^2\Gamma(-\alpha+1)+\mathcal{O}(d_{\e}^{\alpha-1}\lambda^\alpha)\ \label{eq: extended operator} \ .
\end{align}
In~\eqref{eq: extended operator}, $d_{\e}(\mathbf{x}_h) := \tau_0(\mathbf{x}_h) \, \e^{\mu}$, where $\tau_0(\mathbf{x}_h)$ is defined via~\eqref{eq: levy}. 

%Hence, in the case where $\beta_h$ is defined via~\eqref{eq: beta} and~\eqref{eq: levy}, we have the following approximate form of the first term on the right-hand side of transport equation~\eqref{eq:transpeqphresc}
%\begin{equation}
%(\mathds{1}-\mathcal{T}_h)[p^{\beta_h}_{h \e}] = \hat{B}_\varepsilon(\mathbf{x}_h,\varepsilon^{1-\gamma}c\mathbf{v}_h\cdot\nabla_{\mathbf{x}_h})p_{h \e} \label{eq:Tepslevy} \ ,
%\end{equation}
%with $\hat{B}_\varepsilon$ being defined via~\eqref{eq: extended operator}.

\paragraph{Transport equations for $\rho_{I}^\e$, $\rho_{A}^\e$ and $\rho_{D}^\e$} Integrating both sides of transport equation~\eqref{eq:transpeqphresc} with respect to ${\bf v}_h$ over the set ${\rm V}$ and using the fact that the turning operator $\mathcal{T}_h$ satisfies~\eqref{eq: zero turn operator}, we find that the macroscopic cell density $\rho_{h}^\e(\mathbf{x}_h,t)$ given by~\eqref{eq: macroscopic quantities} satisfies the following transport equation 
\begin{align}
   \partial_t \rho_{h}^\e + 2c \, \nabla_{\mathbf{x}_h} \cdot w_{h}^\e &= \e^{-1} \int_{\rm V} \mathcal{Q}_{hk}^\e \diff\mathbf{v}_h \ , \quad \mathbf{x}_h \in \mathbb{R}^n, t \in \mathbb{R}^*_+ \ .
     \label{eq:transpeqrhoepstemp}
\end{align}

Moreover, substituting the expressions for $p_{h}^\e(\mathbf{x}_h,t,\mathbf{v}_h)$ and $p_{k}^\e(\mathbf{x}_h,t,\mathbf{v}_k)$ given by~\eqref{eq:phfact} into the definitions of $\mathcal{K}_{hk}^\e$, $ \mathcal{J}_{hk}^\e$ and ${}_{\e}\mathcal{J}^l_{hk}$ given by~\eqref{def:Jhkeps}-\eqref{def:opmeps} we find
\begin{equation}
 \int_{\rm V} \mathcal{K}_{hk}^\e(\mathbf{x}_h,t,\mathbf{v}_h)  \diff\mathbf{v}_h = 0
  \label{eq:Khkepsfact} 
\end{equation}
and, neglecting higher order terms, we also obtain
\begin{equation}
\int_{\rm V} \mathcal{J}_{hk}^\e(\mathbf{x}_h,t,\mathbf{v}_h) \diff\mathbf{v}_h  = -\e^{\xi -\vartheta - \gamma} \, c \, M \,  \rho_{h}^\e \,  \rho_{k}^\e\ , \quad  \int_{\rm V} {}_{\e}\mathcal{J}^h_{lk}(\mathbf{x}_h,t,\mathbf{v}_h) \diff\mathbf{v}_h  = \e^{\xi -\vartheta - \gamma} \, c \, M \,  \rho_{l}^\e \,  \rho_{k}^\e \ ,
 \label{eq:JMhkepsfact} 
\end{equation}
where $M$ is given by
\begin{equation}
\label{eq: b1app}
M:=\frac{1}{|{\rm V}|^3}\int_{\rm V}\int_{\rm V}\int_{{\rm V}^+_{hk}}\nu\cdot(\mathbf{v}_h-\mathbf{v}_k)\diff\nu\diff\mathbf{v}_h\diff\mathbf{v}_k\ ,\quad h, k \in \{A, D, I \}, \ h \neq k \ .
\end{equation}
Notice that relation~\eqref{eq:Khkepsfact} is obtained using property~\eqref{eq:propKhk}.\\

In conclusion, using~\eqref{eq:Khkepsfact} and~\eqref{eq:JMhkepsfact} along with~\eqref{eq:colliratesIDeps}-\eqref{eq:colliratesDIeps}, from transport equation~\eqref{eq:transpeqrhoepstemp} we obtain the following equations for the macroscopic cell densities $\rho_{I}^\e(\mathbf{x}_I,t)$, $\rho_{A}^\e(\mathbf{x}_A,t)$ and $\rho_{D}^\e(\mathbf{x}_D,t)$:
\begin{align}
    \partial_t\rho_{I}^\e+ 2c \, \nabla_{\mathbf{x}_I}\cdot w_{I}^\e&=- \, c \, M \, \rho_{I}^\e\rho_{D}^\e \ ,&& \mathbf{x}_I \in \mathbb{R}^2, t \in \mathbb{R}^*_+ \ \label{conservation levy},\\[5pt]
    \partial_t\rho_{A}^\e + 2c \, \nabla_{\mathbf{x}_A}\cdot w_{A}^\e &= \, c \, M \, \rho_{I}^\e \rho_{D}^\e \ , && \mathbf{x}_A \in \mathbb{R}^2, t \in \mathbb{R}^*_+ \ ,\label{eq: final conservation brownian}\\[5pt]
    \partial_t\rho_{D}^\e + 2c \, \nabla_{\mathbf{x}_D}\cdot w_{D}^\e&=0  \ , && \mathbf{x}_D \in \mathbb{R}^2, t \in \mathbb{R}^*_+ \ .\label{eq: conservation final cancer}
\end{align}
%{\red From the macroscopic equations~\eqref{conservation levy} and~\eqref{eq: final conservation brownian} we choose $\kappa=-(\xi-\vartheta)+3/2>0$ for $\gamma=1/2$.}
Here,  we have used the scaling relations in~\eqref{ass:scal1} for the parameter $\kappa$.
{ On the right hand side of~\eqref{conservation levy} we have the density of cells that are leaving the state $I$ (due to interactions with cells in the population $D$) and are appearing in the new state $A$ in~\eqref{eq: final conservation brownian}.}

\paragraph{Transport equations for $w_{I}^\e$, $w_{A}^\e$ and $w_{D}^\e$} Multiplying both sides of transport equation~\eqref{eq:transpeqphresc} by ${\bf v}_h$ and then integrating both sides of the resulting equation with respect to ${\bf v}_h$ over the set ${\rm V}$, we find that the local macroscopic direction of cell motion $w_{h}^\e(\mathbf{x}_h,t)$ given by~\eqref{eq: macroscopic quantities} satisfies the following transport equation 
\begin{align}
    \varepsilon^{1+\gamma}2\partial_tw_{h}^\e +\varepsilon^{1-\gamma} c\nabla_{\mathbf{x}_h}\int_{\rm V}\mathbf{v}_h\otimes\mathbf{v}_h \, {p}_{h}^\e\diff\mathbf{v}_h = &- \int_{\rm V} \mathbf{v}_h (\mathds{1}-\mathcal{T}_h)[{}_{\e}p^{\beta_h}_{h}] \diff\mathbf{v}_h  \nonumber \\  & +  \int_{\rm V} \mathbf{v}_h \, \mathcal{Q}_{hk}^\e \diff\mathbf{v}_h \ .
     \label{eq:transpeqweps1}
\end{align}

In the case where $\beta_h$ is defined via~\eqref{eq: beta} and~\eqref{eq: brownian}, using~\eqref{eq:phfact}, \eqref{eq:phbetabro} and the properties of the turning operator $\mathcal{T}_h$ established by Lemma~\ref{lem: eigenfunctions} we find that the first term on the right-hand side of~\eqref{eq:transpeqweps1} is given by
\begin{equation}
\int_{\rm V} \mathbf{v}_h (\mathds{1}-\mathcal{T}_h)[{}_{\e}p^{\beta_h}_{h}] \diff\mathbf{v}_h= \frac{2\varepsilon^\gamma}{|{\rm V}|} b{(1-\iota_1)}w_h^\e \ . \label{eq:Tepsvbrownian}
\end{equation}
Here, $\iota_1(\mathbf{x}_h,t)$ is the first non-zero eigenvalue of the turning operator $\mathcal{T}_h$, which is given by~\eqref{eq: eigen}. On the other hand, when $\beta_h$ is defined via~\eqref{eq: beta} and~\eqref{eq: levy}, using~\eqref{eq:phfact}, \eqref{eq:phbetalev} and the properties of the turning operator $\mathcal{T}_h$ established by Lemma~\ref{lem: eigenfunctions}, it was proved in~\cite{estrada2018fractional} that the following approximate expression of the first term on the right-hand side of~\eqref{eq:transpeqweps1} holds
\begin{align}
\int_{\rm V} &\mathbf{v}_h (\mathds{1}-\mathcal{T}_h)[{}_{\e}p^{\beta_h}_{h}] \diff\mathbf{v}_h = \varepsilon^{1-\frac{\gamma}{\alpha-1}} \Bigl(g_\alpha\nabla^{\alpha-1}\rho^\e_h{-}\frac{2(\alpha-1)}{\tau_0|{\rm V}|} (\iota_1-1) w_h^\e\Bigr) +\textnormal{l.o.t.}
\ , \label{eq:Tepsvlevy}
\end{align}
where
\begin{equation}\label{def:calpha}
g_\alpha(\mathbf{x}_h,t):=\frac{\pi\tau_0^{\alpha-2}(1-\alpha)^2c^{\alpha-1}}{\sin(\pi\alpha)\Gamma(\alpha)}\Bigl(\frac{4\iota_1-|{\rm V}|}{|{\rm V}|}\Bigr)\ \ \textnormal{for}\ \ \Gamma(-\alpha+1)=\frac{\pi}{\sin(\pi\alpha)\Gamma(\alpha)} \ .
\end{equation}
Notice that $g_\alpha(\cdot,\cdot)>0$ since $\sin(\pi\alpha)<0$  for $\alpha\in(1,2)$ and  $4 \iota_1-|{ {\rm V}}|<0$ by using \eqref{volumen} for $n=2$ and recalling that $\iota_1<1$.  \\

Moreover, as similarly done in~\cite{estrada2020interacting}, using the fact that $(\cdot)':{\rm V}\mapsto {\rm V}$ is a bijection and $\mathbf{v}_h'\cdot\nu=-\mathbf{v}_h\cdot\nu$, whence $\nu\cdot(\mathbf{v}_h-\mathbf{v}_k)=-\nu\cdot(\mathbf{v}_h'-\mathbf{v}_k')$, we find
\begin{align}
\int_{\rm V} & \mathbf{v}_h \, \mathcal{K}_{hk}^\e \, \diff\mathbf{v}_h \nonumber \\
& = \frac{1}{|{{\rm V}}|} \e^{\xi -\vartheta - \gamma} \, c \, \Big(\int_{\rm V} \int_{\rm V} \int_{{\rm V}^+_{hk}} {\bf v}_h \, p_{h}^\e(\mathbf{x}_h,t,{\bf v}_h') p_{k}^\e(\mathbf{x}_h,t,{\bf v}_k') \nu\cdot(\mathbf{v}_h-\mathbf{v}_k) \diff\nu \diff\mathbf{v}_k \diff\mathbf{v}_h  \nonumber\\
     &- \int_{\rm V} \int_{\rm V} \int_{{\rm V}^+_{hk}} {\bf v}_h \, p_{h}^\e(\mathbf{x}_h,t,{\bf v}_h) p_{k}^\e(\mathbf{x}_h,t,{\bf v}_k) \nu\cdot(\mathbf{v}_h-\mathbf{v}_k) \, \diff\nu\diff\mathbf{v}_k \diff\mathbf{v}_h \Big) \nonumber\\
&=-  \frac{1}{|{{\rm V}}|} \e^{\xi -\vartheta - \gamma} \, c \, \int_{\rm V}\int_{\rm V}\int_{{\rm V}^+_{hk}}(\mathbf{v}'_h)'p_{h}^\e(\mathbf{x}_h,t,\mathbf{v}_h')p_{k}^\e(\mathbf{x}_h,t,\mathbf{v}_k')\nu\cdot(\mathbf{v}_h'-\mathbf{v}_k')\diff\nu\diff\mathbf{v}_h'\diff\mathbf{v}_k'\nonumber\\ &\ \ \ -\int_{\rm V}\int_{\rm V}\int_{{\rm V}^+_{hk}}\mathbf{v}_h p_{h}^\e(\mathbf{x}_h,t,\mathbf{v}_h)p_{k}^\e(\mathbf{x}_h,t,\mathbf{v}_k)\nu\cdot(\mathbf{v}_h-\mathbf{v}_k)\diff\nu\diff\mathbf{v}_h\diff\mathbf{v}_k\nonumber\\ &=  \frac{1}{|{{\rm V}}|} \e^{\xi -\vartheta - \gamma} \, c \, \int_{\rm V}\int_{\rm V}\int_{{\rm V}^+_{hk}}(\mathbf{v}_h'-\mathbf{v}_h)p_{h}^\e(\mathbf{x}_h,t,\mathbf{v}_h)p_{k}^\e(\mathbf{x}_h,t,\mathbf{v}_k)\nu\cdot(\mathbf{v}_h-\mathbf{v}_k)\diff\nu\diff\mathbf{v}_h\diff\mathbf{v}_k\nonumber\\ & =-  \frac{1}{|{{\rm V}}|} \e^{\xi -\vartheta - \gamma} \, c \, \frac{4}{3}\int_{\rm V}\int_{\rm V}|\mathbf{v}_h-\mathbf{v}_k| \, \mathbf{v}_h \, p_{h}^\e(\mathbf{x}_h,t,\mathbf{v}_h) \, p_{k}^\e(\mathbf{x}_h,t,\mathbf{v}_k)\diff\mathbf{v}_h\diff\mathbf{v}_k\ . \label{eq: long computation}
\end{align}
The last equality in~\eqref{eq: long computation} is obtained using the fact that $\mathbf{v}_h'-\mathbf{v}_h=-2(\mathbf{v}_h\cdot\nu)\nu$. Substituting the expressions for $p_{h}^\e(\mathbf{x}_h,t,\mathbf{v}_h)$ and $p_{k}^\e(\mathbf{x}_h,t,\mathbf{v}_k)$ given by~\eqref{eq:phfact} into~\eqref{eq: long computation} and into definitions~\eqref{def:Jhtoeps} and \eqref{def:opmeps} of $ \mathcal{J}_{hk}^\e$ and ${}_{\e}\mathcal{J}^h_{lk}$, neglecting higher order terms we find
\begin{equation}
 \int_{\rm V} \mathbf{v}_h \, \mathcal{K}_{hk}^\e  \diff\mathbf{v}_h =  -  \e^{\xi -\vartheta} \, c \, \frac{8}{3} \, \dfrac{1}{|{\rm V}|^3} \, q_h \, \rho_k^\e w_h^\e\ ,
  \label{eq:Khkepsfactw} 
\end{equation}
and
{\begin{equation}
\int_{\rm V} \mathbf{v}_h \, \mathcal{J}_{hk}^\e \diff\mathbf{v}_h  = - \frac{\e^{\xi -\vartheta}2c}{|{\rm V}|^2}  \tilde{q}_h\rho_k^\e w_h^\e \ , \quad  \int_{\rm V} \mathbf{v}_h \, {}_{\e}\mathcal{J}^h_{lk} \diff\mathbf{v}_h  =  \frac{\e^{\xi -\vartheta}2c}{|{\rm V}|^2}   \tilde{q}_h\rho_k^\e w_h^\e \ ,
 \label{eq:JMhkepsfactw} 
\end{equation}}
where  $q_h$ and $\tilde{q}_h$ are defined as
\begin{equation}
\label{def:smallqh}
q_h:=\int_{{\rm V}}|\mathbf{v}_h-\mathbf{v}_k|\diff\mathbf{v} \ , \quad { \tilde{q}_h:=\int_{{\rm V}^+_{hk}}|\nu||\mathbf{v}_h-\mathbf{v}_k|\diff \nu}\ , \quad h \in \{A, D, I\} \ . 
\end{equation}

Finally, substituting the expression of $p_{h}^\e(\mathbf{x}_h,t,\mathbf{v}_h)$ into the second term on the left-hand side of~\eqref{eq:transpeqweps1} and neglecting higher order terms yields
\begin{equation}
c \, \nabla_{\mathbf{x}_h}\int_{\rm V}\mathbf{v}_h\otimes\mathbf{v}_h \, {p}_{h}^\e \diff\mathbf{v}_h = C_h \, \nabla_{\mathbf{x}_h} {\rho}_{h}^\e \ ,
 \label{eq:Jtransplefteps} 
\end{equation}
with $C_h$ being defined as
\begin{equation}\label{def:ch}
C_h := \dfrac{c}{|{\rm V}|}\int_{\rm V}\mathbf{v}_h\otimes\mathbf{v}_h\diff\mathbf{v}_h \ , \quad h \in \{A, D, I\} \ .
\end{equation}

In conclusion, using~\eqref{eq:phbetabro} and \eqref{eq:phbetalev}, \eqref{eq:Khkepsfactw}, \eqref{eq:JMhkepsfactw} and~\eqref{eq:Jtransplefteps} along with~\eqref{eq:colliratesIDeps}-\eqref{eq:colliratesDIeps}, from transport equation~\eqref{eq:transpeqweps1} we obtain the following transport equations for the local macroscopic directions of cell motion $w_{I}^\e(\mathbf{x}_I,t)$, $w_{A}^\e(\mathbf{x}_A,t)$ and $w_{D}^\e(\mathbf{x}_D,t)$:

\begin{align}
    &\varepsilon^{1+\gamma}2\partial_tw_{I}^\e+\varepsilon^{1-\gamma} C_I \, \nabla_{\mathbf{x}_I} \rho_{I}^\e ={ -} \varepsilon^{1-\frac{\gamma}{\alpha-1}}\left(g_\alpha\nabla^{\alpha-1}\rho_{I}^\e{ -}\frac{2(\alpha-1)}{\tau_0|{\rm V}|} (\iota_1-1) w_{I}^\e\right)\nonumber\\ & - \varepsilon^{\upxi-\vartheta} \, c \,  \left((1-\e^{\kappa})\frac{8}{3}\frac{1}{|{\rm V}|^3} q_I \, \rho_{D}^\e w_{I}^\e +{ \frac{2c}{|{\rm V}|^2}\tilde{q}_I\rho_D^\e w_I^\e}\right)  \ , \quad \mathbf{x}_I \in \mathbb{R}^2, t \in \mathbb{R}^*_+ \ , \label{eq: levy mean direction}
\end{align}

\begin{align}
    \varepsilon^{1+\gamma}2\partial_tw_{A}^\e+\varepsilon^{1-\gamma} &C_A \, \nabla_{\mathbf{x}_A} \rho_{A}^\e =  -\frac{2\varepsilon^\gamma}{|{\rm V}|} b{ (1-\iota_1)}w_{A}^\e \nonumber\\& -  \varepsilon^{\upxi-\vartheta} \, c \, \left(\frac{8}{3}\frac{1}{|{\rm V}|^3} q_A \, \rho_{D}^\e w_{A}^\e- {\frac{2c}{|{\rm V}|^2}\tilde{q}_A\rho_D^\e w_A^\e}\right)  \ , \quad \mathbf{x}_A \in \mathbb{R}^2, t \in \mathbb{R}^*_+ \ , \label{eq: brownian mean direction}
%    \qquad\qquad+ \varepsilon^{\upxi-\vartheta} \left(\zeta I_{ID} \rho_{D \e} w_{I \e} - \frac{4}{3}\frac{c}{|{\rm V}|^3} q_A \, \rho_{D \e} w_{A \e}\right) \ ,\label{eq: brownian mean direction}
\end{align}
and
\begin{align}
    \varepsilon^{\gamma+1}2\partial_t w_{D}^\e +\varepsilon^{1-\gamma} C_D \, \nabla_{\mathbf{x}_D} & \rho_{D}^\e = -\frac{2\varepsilon^{\gamma}}{|{\rm V}|} b(\iota_1-1)w_{D}^\e\nonumber\\ & - \varepsilon^{\upxi-\vartheta} \, c \, \frac{8}{3} \frac{1}{|{\rm V}|^3} q_D \, \rho_{I}^\e w_{D}^\e  \ , \quad \mathbf{x}_D \in \mathbb{R}^2, t \in \mathbb{R}^*_+ \ .\label{eq: boring}
\end{align}

\paragraph{Macroscopic scale model} Noting that $1-\dfrac{\gamma}{\alpha-1} < 1 -\gamma$ since $\alpha \in (1,2)$ and using assumptions~\eqref{ass:scal1} { for the scaling parameters}, letting $\e \to 0$ in~\eqref{eq: levy mean direction}-\eqref{eq: boring} we formally find the following expressions for the leading-order terms $w_I({\bf x},t)$, $w_A({\bf x},t)$ and $w_D({\bf x},t)$ of the asymptotic expansions for the local macroscopic directions of cell motion $w_{I \e}({\bf x},t)$, $w_{A \e}({\bf x},t)$ and $w_{D \e}({\bf x},t)$:
\begin{itemize}
\item [-] From \eqref{eq: levy mean direction} and choosing the scaling parameters as $1-\frac{\gamma}{\alpha-1}=\xi-\vartheta$ we have
\begin{equation}
    w_I=-\frac{g_{\alpha}}{H_I}\nabla_{\mathbf{x}}^{\alpha-1}\rho_I\ , \quad \textnormal{where} \quad H_I :=\frac{2(\alpha-1)(1-\iota_1)}{\tau_0|{\rm V}|}+\frac{8c}{|3{\rm V}|^3}q_I\rho_D>0\ .
    \label{mean direction 1}
\end{equation}
\item [-] From \eqref{eq: brownian mean direction} and choosing $\xi-\vartheta=\gamma$ in agreement with \eqref{ass:scal1}, we obtain
\begin{equation}
    w_A=-\frac{C_A}{H_A}\nabla_{\mathbf{x}}\rho_A\ , \quad \textnormal{where}\quad H_A :=\frac{2b}{|{\rm V}|}(1-\iota_1)+\frac{8c}{3|{\rm V}|^3}q_A\rho_D>0\ .
    \label{mean direction 2}
\end{equation}
\item [-] Finally, using the same scaling rules as in the previous case we obtain from \eqref{eq: boring}
\begin{equation}
    w_D=-\frac{C_D}{H_D}\nabla_{\mathbf{x}}\rho_D\ ,\quad \textnormal{where}\quad H_D :=\frac{2b}{|{\rm V}|}(1-\iota_1)+\frac{8c}{3|{\rm V}|^3}q_D\rho_I>0\ .\label{mean direction 3}
\end{equation}
\end{itemize}
\iffalse
\begin{equation}
w_I=-\frac{g_\alpha}{\sigma_{\alpha}+\sigma_{q_I}\rho_D}\nabla^{\alpha-1}_{\mathbf{x}}\rho_I\ , \quad  w_A =-\frac{C_A}{\sigma_{b}-\sigma_{q_A}\rho_D}\nabla_{\mathbf{x}}\rho_A, \quad w_D=-\frac{C_D}{\sigma_{b}-\sigma_{q_D}\rho_I}\nabla_{\mathbf{x}}\rho_D
\label{eq: mean directions}
\end{equation}
with
\begin{equation}
\begin{aligned}
\label{def:sigmaalphab}
 \sigma_{\alpha}({\bf x},t) &:=\frac{2(\alpha-1)(\iota_1({\bf x},t)-1)}{\tau_0({\bf x})| {\rm V}|}\ , \quad \sigma_{b}({\bf x},t) := \frac{2b}{|{\rm V}|}(\iota_1({\bf x},t)-1)\ ,\\
  \quad \sigma_{q_h}&:=-\frac{c}{|V|^3}\frac{8}{3}q_h\ , \quad \textnormal{for}\quad h \in \{I,A,D\}\ .
\end{aligned}
\end{equation}
\fi
Furthermore, under assumptions~\eqref{ass:scal1}, letting $\e \to 0$ in~\eqref{conservation levy}-\eqref{eq: conservation final cancer} and using\eqref{mean direction 1}-\eqref{mean direction 3}, we formally obtain the following balance equations for the leading-order terms $\rho_I({\bf x},t)$, $\rho_A({\bf x},t)$ and $\rho_D({\bf x},t)$ of the asymptotic expansions for the macroscopic cell densities $\rho_{I}^\e({\bf x},t)$, $\rho_{A}^\e({\bf x},t)$ and $\rho_{D}^\e({\bf x},t)$
 \begin{align}
    \partial_t\rho_I-\nabla_{\mathbf{x}}\cdot\Bigl(D_I \, \nabla^{\alpha-1}_{\mathbf{x}}\rho_I \Bigr)&=- a \, \rho_I \,\rho_D \ , \quad \alpha \in (1,2) \ , &&  \mathbf{x} \in \mathbb{R}^2, t \in \mathbb{R}^*_+\ ,\label{eq: final I}\\
    \partial_t\rho_A-\nabla_{\mathbf{x}}\cdot\Bigl(D_A \, \nabla_{\mathbf{x}}\rho_A \Bigr)&= a \, \rho_I \,\rho_D \ , && \mathbf{x} \in \mathbb{R}^2, t \in \mathbb{R}^*_+ \ ,\label{eq: final A}\\
     \partial_t\rho_D-\nabla_{\mathbf{x}}\cdot \Bigl(D_D \, \nabla_{\mathbf{x}}\rho_D\Bigr)&=0\ ,  && \mathbf{x} \in \mathbb{R}^2, t \in \mathbb{R}^*_+ \ ,\label{eq: final D}
 \end{align}
% \begin{align}
%    \partial_t\rho_I-2c\nabla_{\mathbf{x}}\cdot\Bigl(\frac{g_\alpha}{\sigma_{\alpha}+\sigma_{q_I}\rho_D}\nabla^{\alpha-1}_{\mathbf{x}}\rho_I \Bigr)&=- c \, M \, \rho_I \,\rho_D \ , \quad \alpha \in (1,2) \ , &&  \mathbf{x} \in \mathbb{R}^2, t \in \mathbb{R}^*_+\ ,\label{eq: final I}\\
%    \partial_t\rho_A-2c\nabla_{\mathbf{x}}\cdot\Bigl(\frac{C_A}{\sigma_{b}-\sigma_{q_A}\rho_D}\nabla_{\mathbf{x}}\rho_A \Bigr)&= c \, M \, \rho_I \,\rho_D \ , && \mathbf{x} \in \mathbb{R}^2, t \in \mathbb{R}^*_+ \ ,\label{eq: final A}\\
%     \partial_t\rho_D-2c\nabla_{\mathbf{x}}\cdot \Bigl(\frac{C_D}{\sigma_{b}-\sigma_{q_D}\rho_I}\nabla_{\mathbf{x}}\rho_D\Bigr)&=0\ ,  && \mathbf{x} \in \mathbb{R}^2, t \in \mathbb{R}^*_+ \ ,\label{eq: final D}
% \end{align}
where 
$$
D_I := \frac{2 \, c \, g_\alpha}{H_I}, \quad D_A := \frac{2 \, c \, C_A}{H_A}, \quad D_D := \frac{2 \, c \, C_D}{H_D}, \quad a := c \, M. 
$$
%and $M$ is defined via~\eqref{eq: b1app}.  

\begin{rem}
Notice that the  the functions $D_I$, $D_A$ and $D_D$ are strictly positive. Moreover, the dependence of { these functions} on the cell densities follows from conservative interactions between cells of different populations, while population-switching interactions do not affect { their values}.
\end{rem}

\paragraph{ Considerations on the macroscopic scale model { \eqref{eq: final I}-\eqref{eq: final D}}} The functions $\rho_I({\bf x},t)$, $\rho_A({\bf x},t)$ and $\rho_D({\bf x},t)$ model, respectively, the density of CTLs in the pre-activation state, activated CTLs and DCs presenting a tumour antigen on their surface at position ${\bf x}$ and time $t$. The spatio-temporal coevolution of CTLs and DCs is modelled through the coupled system of balance equations~\eqref{eq: final I}-\eqref{eq: final D}, which governs the dynamics of the cell density functions. 

{ The mathematical model defined by~\eqref{eq: final I}-\eqref{eq: final D} provides a macroscopic description of cell dynamics that takes explicitly into account the effects of cell-cell interactions and the characteristics of cell motion that are encapsulated in the parameters $c$ (i.e. the magnitude of the cell velocity, which is assumed to be constant), $\alpha$ (i.e. the characteristic exponent of the long-tailed distribution followed by the running time of CTLs in the pre-activation state) and $b$ (i.e. the characteristic exponent of the Poisson distribution followed by the running time of activated CTLs and DCs).}

This model effectively captures the fact that interactions between DCs presenting a tumour antigen on their surface and CTLs in the pre-activation state lead to CTL activation. In particular, the term on the right-hand side of~\eqref{eq: final I} models the decay in the density of CTLs in the pre-activation state at position ${\bf x}$ and time $t$ due to contact interactions with DCs which result in CTL activation, while the term on the right-hand side of~\eqref{eq: final A} models the corresponding growth in the density of activated CTLs. As one would expect, these two terms differ only in their signs and are proportional to the product between the cell density functions $\rho_{I}({\bf x},t)$ and $\rho_{D}({\bf x},t)$. The factor of proportionality $a$ increases with the value of the parameter $c$. This is coherent with the observation that higher cell motilities may increase the encounter rate of CTLs in the pre-activation state with DCs.

The model captures also the fact that CTLs in the pre-activation state move in a non-local search pattern, while the search pattern of activated CTLs is more localised. In fact, the rate of change of the density of CTLs in the pre-activation state due to cell movement (i.e. the second term on the left-hand side of~\eqref{eq: final I}) is a fractional diffusion term, while that of the density of activated CTLs (i.e. the second term on the left-hand side of~\eqref{eq: final A}) is a classical diffusion term. The function modelling the diffusivity of CTLs in the pre-activation state (i.e. the function $D_I$) and the function modelling the diffusivity of activated CTLs (i.e. the function $D_A$) are proportional to the parameter $c$. { This is coherent with the observation that, {\it ceteris paribus}, a higher magnitude of the cell velocity correlates with a higher cell motility. Both $D_I$ and $D_A$ are monotonically decreasing functions of the cell density function $\rho_{D}({\bf x},t)$, which means that, all else being equal, the higher the density of DCs at a given position, the lower the diffusivity of CTLs. This reflects the fact that higher densities of DCs will make it more likely that interactions between CTLs and DCs occur and, since these interactions force CTLs to change their direction of movement at the mesoscopic scale, this will ultimately result in a lower cell diffusivity at the macroscopic scale. Moreover, $D_A$ is an increasing function of $b$. This is coherent with the fact that larger values of this parameter correspond to larger mean values of the cell running times.}

The fact that the right-hand side of~\eqref{eq: final D} is zero translates in mathematical terms the idea that we are not taking into account the effects of division and death of DCs. Moreover, coherently with the fact that the motion of DCs is here described as a Brownian motion, the rate of change of the density of DCs due to cell movement (i.e. the second term on the left-hand side of~\eqref{eq: final D}) is a classical diffusion term. Considerations analogous to those made above about the dependence of the function $D_A$ on the parameters $c$ and $b$ apply to the function modelling the diffusivity of DCs (i.e. the function $D_D$) as well. Furthermore, considerations similar to those made above about the dependence of $D_I$ on the density function $\rho_{D}({\bf x},t)$ hold for the dependence of $D_D$ on the density function $\rho_{I}({\bf x},t)$.

\section{Research perspectives}
\label{Sec:persp}
The modelling approach for the switch between cell migration modes presented here could be generalised by including additional cellular phenomena involved in the immune response to cancer, and considering other aspects of immune cell movement as well. { For the case of movement in bacteria,} a recent work in this direction is \cite{perthame2018fractional}, where { the switch in type of movement}, i.e. the switch between L\'{e}vy and Brownian strategies, was determined by  chemical pathways internal to the bacteria.

With reference to the mathematical modelling of the immune response to cancer, a natural generalisation would be to include a population of cancer cells and allow activated CTLs to induce death in cancer cells via binary interactions. Moreover, the recognition phase of the adaptive immune response to cancer could be modelled by splitting the population of DCs into a subpopulation of cells with no tumour antigens on their surface and a subpopulation of cells presenting some antigen -- which would move in a non-local and in a more localised search pattern, respectively~\cite{engelhardt2012marginating} -- and  letting DCs switch from one subpopulation to the other via binary interactions with cancer cells  ~\cite{waldman2020guide,wculek2019dendritic}. The strategy we have used here to model non-conservative cell-cell interactions may prove useful to the development of both generalisations of our modelling approach. 

In regard to the mathematical modelling of other aspects of immune cell movement, our modelling approach could be extended to represent other switches in T cell migration patterns observed in the immune response to different pathogens, which are driven by possible chemotactic cues and by the conditions of the surrounding microenvironment~\cite{krummel2016t}. Moreover, further generalisations of the modelling approach could be developed in relation to experimental results indicating that T cells can also undergo subdiffusive~\cite{worbs2007ccr7} and fully ballistic~\cite{witt2005directed} migration. 

In general, it would be interesting to apply the modelling approach presented in this paper and its possible developments to other biological and ecological contexts whereby switch from non-local to localised migration patterns has been reported~\cite{bartumeus2003helical,de2014superdiffusion,de2007patch,humphries2010environmental,nolet2002search}.

We conclude by remarking that, as previously noted, although they may result in cell outgoing trajectories compatible with those observed in elastic collisions, binary collisions between cells are not elastic in nature. Hence, it will be necessary to go beyond the definition of post-collision velocities used here in order to have a more biophysically faithful representation of cell-cell interactions. This is beyond the scope of the present work, which is primarily focused on modelling the switch in T cell migration modes mediated by interactions between inactive CTLs and DCs. Moreover, the formal approach employed in this article to derive a macroscopic limit of the mesoscopic model relies on the assumption that cell densities are sufficiently low so that cell velocities can be assumed to be uncorrelated. As such, it may lead to an inaccurate mean field representation of the dynamics of the underlying biological system in cases where cell densities are not sufficiently low, or cell-cell interactions introduce a stronger correlation between cell velocities. Therefore, another fruitful avenue of research would lie in extending this formal approach to these more complex cases by identifying alternative ways of obtaining a closed system of coupled equations for the macroscopic cell densities starting from the corresponding kinetic model.

\appendix

\section{Derivation of transport equations~\eqref{eq: 13 equation}-\eqref{eq: 13 equationg}}\label{sec: appendix A}
Using the method presented in~\cite{estrada2020interacting}, we show how to derive a transport equation for the two-particle distribution function $\tilde{\tilde{f}}_{hk}(\mathbf{x}_h,\mathbf{x}_k,t,\mathbf{v}_h,\mathbf{v}_k)$ starting from transport equation~\eqref{eq: transpfhkgen} for the two-particle distribution function $f_{hk}(\mathbf{x}_h,\mathbf{x}_k,t,\mathbf{v}_h,\mathbf{v}_k,\tau_h,\tau_k)$.

We first introduce the notation
$$
\tilde{f}_{\tau_h}(\mathbf{x}_h,\mathbf{x}_k,t,\mathbf{v}_h,\mathbf{v}_k, \tau_h) := \int_0^tf_{hk}(\mathbf{x}_h,\mathbf{x}_k,t,\mathbf{v}_h,\mathbf{v}_k,\tau_h,\tau_k) \diff\tau_k\ ,
$$
and
$$
\tilde{f}_{\tau_k}(\mathbf{x}_h,\mathbf{x}_k,t,\mathbf{v}_h,\mathbf{v}_k, \tau_k) := \int_0^tf_{hk}(\mathbf{x}_h,\mathbf{x}_k,t,\mathbf{v}_h,\mathbf{v}_k,\tau_h,\tau_k) \diff\tau_h\ ,
$$
and then note that, when $\beta_h$ and $\beta_k$ are given by~\eqref{eq: beta} with~$\psi_h$ and~$\psi_k$ defined via~\eqref{eq: brownian} or~\eqref{eq: levy}, the solutions of~\eqref{eq: transpfhkgen} subject to the initial and boundary conditions considered here are such that $\tilde{f}_{\tau_h}$ decays monotonically as $\tau_h$ increases, and $\tilde{f}_{\tau_k}$ exhibits an analogous behaviour. Hence, integrating~\eqref{eq: transpfhkgen} with respect to $(\tau_h, \tau_k)$ over $(0,t)^2$ with $t$ large enough so that $\tilde{f}_{\tau_h}(\mathbf{x}_h,\mathbf{x}_k,t,\mathbf{v}_h,\mathbf{v}_k, \tau_h=t)$ is negligible compared to $\tilde{f}^0_{\tau_h}(\mathbf{x}_h,\mathbf{x}_k,t,\mathbf{v}_h,\mathbf{v}_k)$ and $\tilde{f}_{\tau_k}(\mathbf{x}_h,\mathbf{x}_k,t,\mathbf{v}_h,\mathbf{v}_k, \tau_k=t)$ is negligible compared to $\tilde{f}_{\tau_k}^0(\mathbf{x}_h,\mathbf{x}_k,t,\mathbf{v}_h,\mathbf{v}_k)$, with
$$
\tilde{f}^0_{\tau_h}(\mathbf{x}_h,\mathbf{x}_k,t,\mathbf{v}_h,\mathbf{v}_k) := \tilde{f}_{\tau_h}(\mathbf{x}_h,\mathbf{x}_k,t,\mathbf{v}_h,\mathbf{v}_k, \tau_h=0)
$$
and
$$
\tilde{f}_{\tau_k}^0(\mathbf{x}_h,\mathbf{x}_k,t,\mathbf{v}_h,\mathbf{v}_k) := \tilde{f}_{\tau_k}(\mathbf{x}_h,\mathbf{x}_k,t,\mathbf{v}_h,\mathbf{v}_k, \tau_k=0) \ ,
$$
we obtain the following transport equation for $\tilde{\tilde{f}}_{hk}(\mathbf{x}_h,\mathbf{x}_k,t,\mathbf{v}_h,\mathbf{v}_k)$
$$
    (\partial_t+c \, \mathbf{v}_h\cdot\nabla_{\mathbf{x}_h}+c \, \mathbf{v}_k\cdot\nabla_{\mathbf{x}_k})\tilde{\tilde{f}}_{hk} =-\int_0^t\int_0^t(\beta_h+\beta_k)f_{hk}\diff\tau_h \diff\tau_k +\tilde{f}^0_{\tau_h}+\tilde{f}^0_{\tau_k}\ ,
$$
which can be rewritten as
\begin{equation}
    (\partial_t+c \, \mathbf{v}_h\cdot\nabla_{\mathbf{x}_h}+c \, \mathbf{v}_k\cdot\nabla_{\mathbf{x}_k})\tilde{\tilde{f}}_{hk} = -\tilde{\tilde{f}}^{\beta_h}_{hk} - \tilde{\tilde{f}}^{\beta_k}_{hk} +\tilde{f}^0_{\tau_h}+\tilde{f}^0_{\tau_k}\ ,\label{eq: ij interaction}
\end{equation}
with $\tilde{\tilde{f}}^{\beta_h}_{hk}(\mathbf{x}_h,\mathbf{x}_k,t,\mathbf{v}_h,\mathbf{v}_k)$ and $\tilde{\tilde{f}}^{\beta_k}_{hk}(\mathbf{x}_h,\mathbf{x}_k,t,\mathbf{v}_h,\mathbf{v}_k)$ given by~\eqref{eq: kinetic13ijbeta}. 

When cell movement at the microscopic scale obeys the rules presented in Section~\ref{Sec:micromodel}, we have
\begin{equation}
\begin{aligned}
\tilde{f}^0_{\tau_h} =\mathcal{T}_h[\tilde{\tilde{f}}^{\beta_h}_{hk}] \quad \text{and} \quad \tilde{f}^0_{\tau_k} = \mathcal{T}_k[\tilde{\tilde{f}}^{\beta_k}_{hk}]\ ,\label{eq: new run}
\end{aligned}
\end{equation}
with the turning operators $\mathcal{T}_h$ and $\mathcal{T}_k$ being defined via~\eqref{eq: turn angle operator}. {The first two terms on the right-hand side of~\eqref{eq: ij interaction} describe the density of cells that stop with rates $\beta_h$, $\beta_k$. The initial conditions at $\tau_{h}=0$ and $\tau_k=0$ (i.e. at the beginning of a new run phase) given by~\eqref{eq: new run} describes how the cells will resume their motion in a new direction dictated by the turning operators $\mathcal{T}_h$ and $\mathcal{T}_k$, respectively.} 

Substituting the expressions for $\tilde{f}^0_{\tau_h}$ and  $\tilde{f}^0_{\tau_k}$ given by~\eqref{eq: new run} into transport equation~\eqref{eq: ij interaction} yields 
\begin{align}
    (\partial_t&+c \, \mathbf{v}_h\cdot\nabla_{\mathbf{x}_h} +c \, \mathbf{v}_k\cdot\nabla_{\mathbf{x}_k})\tilde{\tilde{f}}_{hk} =-(\mathds{1}-\mathcal{T}_h)[\tilde{\tilde{f}}^{\beta_h}_{hk}] \nonumber\\ &\quad\quad\quad\quad\quad\quad-(\mathds{1}-\mathcal{T}_k)[\tilde{\tilde{f}}^{\beta_k}_{hk}] \ , \quad ({\bf x}_h,{\bf x}_k) \in \Omega^{2}, t \in \mathbb{R}_+, ({\bf v}_h,{\bf v}_k) \in {\rm V}^2 \ .\label{eq: ij equation}
\end{align}

\begin{rem}
\label{CompSup}
Since we consider transport equation~\eqref{eq: transpfhkgen} complemented with a smooth, compactly supported initial condition, the initial condition for transport equation~\eqref{eq: ij equation} will be a smooth, compactly supported function as well. Therefore, the two-particle distribution function $\tilde{\tilde{f}}_{hk}(\mathbf{x}_h,\mathbf{x}_k,t,\mathbf{v}_h,\mathbf{v}_k)$ will have compact support on $\Omega^{2} \times {\rm V}^{2}$ for all $t \in \mathbb{R}^*_+$.
\end{rem}

\section{Derivation of {the equation for the one-particle distribution}}
\label{sec:appendixB}
Transport equation~\eqref{eq:transpeqph} for the one-particle distribution function $p_h(\mathbf{x}_h,t,\mathbf{v}_h)$ can derived from transport equation~\eqref{eq: ij equation} for the two-particle distribution function $\tilde{\tilde{f}}_{hk}(\mathbf{x}_h,\mathbf{x}_k,t,\mathbf{v}_h,\mathbf{v}_k)$ in six steps as previously done in~\cite{estrada2020interacting}.
\\\\
\noindent {\bf (I)} We integrate transport equation~\eqref{eq: ij equation} with respect to $(\mathbf{x}_k,\mathbf{v}_k)$ over the set $\Omega_k(\mathbf{x}_h)\times {\rm V}$ and multiply both sides of the resulting equation by $|{\rm V}|^{-1}$ to obtain
\begin{align}
 |{\rm V}|^{-1} \int_{\Omega_k(\mathbf{x}_h)}\int_{\rm V}  (\partial_t&+c \, \mathbf{v}_h\cdot\nabla_{\mathbf{x}_h} +c \, \mathbf{v}_k\cdot\nabla_{\mathbf{x}_k})\tilde{\tilde{f}}_{hk} \diff\mathbf{v}_k\diff\mathbf{x}_k = \nonumber\\ & - |{\rm V}|^{-1} \int_{\Omega_k(\mathbf{x}_h)}\int_{\rm V} (\mathds{1}-\mathcal{T}_h)[\tilde{\tilde{f}}^{\beta_h}_{hk}] \diff\mathbf{v}_k\diff\mathbf{x}_k  \nonumber\\ & - |{\rm V}|^{-1}\int_{\Omega_k(\mathbf{x}_h)}\int_{\rm V} (\mathds{1}-\mathcal{T}_k)[\tilde{\tilde{f}}^{\beta_k}_{hk}] \diff\mathbf{v}_k\diff\mathbf{x}_k \ .\label{eq: ij equationint}
\end{align}
\noindent {\bf (II)} Using the fact that $p_h$ is given by~\eqref{eq: p densityhk} and integrals with respect to $\mathbf{x}_k$ and $\mathbf{v}_k$ commute, we rewrite the first term on the left-hand side of~\eqref{eq: ij equationint} as
 \begin{equation*}
        |{\rm V}|^{-1} \partial_t\int_{\Omega_k(\mathbf{x}_h)}\int_{\rm V}\tilde{\tilde{f}}_{hk}\diff\mathbf{v}_k \diff\mathbf{x}_k =\partial_tp_h\ .
\end{equation*}
    
\noindent {\bf (III)} Using Reynold's transport theorem in the variable $\mathbf{x}_h$, we rewrite the second term on the left-hand side of~\eqref{eq: ij equationint} as
\begin{align*}
    |{\rm V}|^{-1} c\int_{\Omega_k(\mathbf{x}_h)}\int_{\rm V}(\mathbf{v}_h\cdot\nabla_{\mathbf{x}_h})&\tilde{\tilde{f}}_{hk}\diff\mathbf{v}_k\diff\mathbf{x}_k=|{\rm V}|^{-1}  c \, \mathbf{v}_h\cdot\nabla_{\mathbf{x}_h}p_h\nonumber\\ &- |{\rm V}|^{-1} c\int_{\partial {\rm B}_\varrho(\mathbf{x}_h)}\int_{\rm V}(\mathbf{v}_h \cdot \nu)\tilde{\tilde{f}}_{hk}\diff\mathbf{v}_k\diff \sigma\ .
\end{align*}
Here, $\nu$ is the unit normal to $\partial \Omega_k(\mathbf{x}_h)$ that points outward from $\Omega_k(\mathbf{x}_h)$ and inward to ${\rm B}_\varrho(\mathbf{x}_i)$, and $\diff \sigma$ denotes the surface element.
\\\\
\noindent {\bf (IV)} Since $\tilde{\tilde{f}}_{hk}$ has compact support on $\Omega^{2} \times \mathbb{\rm V}^{2}$ (\emph{vid.} Remark~\ref{CompSup}), we use the divergence theorem and rewrite the third term on the left-hand side of~\eqref{eq: ij equationint} as
\begin{equation*}
 |{\rm V}|^{-1} c\int_{\Omega_k(\mathbf{x}_h)}\int_{\rm V}(\mathbf{v}_k\cdot\nabla_{\mathbf{x}_k})\tilde{\tilde{f}}_{hk}\diff\mathbf{v}_k\diff\mathbf{x}_k = |{\rm V}|^{-1} c\int_{\partial {\rm B}_\varrho(\mathbf{x}_h)}\int_{\rm V}(\mathbf{v}_k \cdot \nu)\tilde{\tilde{f}}_{hk}\diff\mathbf{v}_k\diff \sigma \ .
\end{equation*}
\\
\noindent {\bf (V)} Changing order of integration, we rewrite the first term on the right-hand side of~\eqref{eq: ij equationint} as 
\begin{equation*}
  -  |{\rm V}|^{-1}\int_{\Omega_k(\mathbf{x}_h)}\int_{\rm V}(\mathds{1}-\mathcal{T}_h)[\tilde{\tilde{f}}^{\beta_h}_{hk}]\diff\mathbf{v}_k\diff\mathbf{x}_k=-(\mathds{1}-\mathcal{T}_h)[p^{\beta_h}_{h}] \ ,
\end{equation*}
with $p^{\beta_h}_{h}(\mathbf{x}_h,t,\mathbf{v}_h)$ given by~\eqref{eq: new densityhkbeta}.
\\\\
\noindent {\bf (VI)} Since $\mathcal{T}_k$ satisfies~\eqref{eq: zero turn operator}, the second term on the right-hand side of~\eqref{eq: ij equationint} is identically zero.
\\

Taken together, the results obtained in Steps {\bf (I)}-{\bf (VI)} allow one to conclude that the one-particle distribution function $p_h(\mathbf{x}_h,t,\mathbf{v}_h)$ satisfies the following transport equation 
\begin{equation}
    \partial_tp_h+c \, \mathbf{v}_h\cdot\nabla_{\mathbf{x}_h}p_h=-(\mathds{1}-\mathcal{T}_h)[p^{\beta_h}_{h}] + \mathcal{Q}_{hk}, \quad \mathbf{x}_h \in \mathbb{R}^n, t \in \mathbb{R}_+, \mathbf{v}_h \in {\rm V} \ ,
\end{equation}
with the weighted one-particle distribution function $p^{\beta_h}_{h}(\mathbf{x}_h,t,\mathbf{v}_h)$ being given by~\eqref{eq: new densityhkbeta} and the term $\mathcal{Q}_{hk}(\mathbf{x}_h,t,\mathbf{v}_h)$ being defined according to~\eqref{def:Qhk}.

\section{Derivation of the non-local trajectory term} 
\label{sec:appendixC}
In the case where $\beta_h$ is defined via~\eqref{eq: beta} and~\eqref{eq: levy} (i.e. for $h=I$) and $\beta_k$ is defined via~\eqref{eq: beta} and~\eqref{eq: brownian} (i.e. for $k=D$), applying the method of characteristics to~\eqref{eq: transpfhkgen} and using the fact that $\dfrac{\psi_k(\cdot,\tau_k)}{\psi_k(\cdot,\tau_k-\tau_h)} = e^{-b\tau_h}$ one finds~\cite{estrada2020interacting}  
\begin{equation}
    f_{hk}=f_{hk}(\mathbf{x}_h-c \, \mathbf{v}_h\tau_h,\mathbf{x}_k-c\mathbf{v}_k\tau_h,t-\tau_h,\mathbf{v}_h,\mathbf{v}_k,\tau_h=0,\tau_k-\tau_h) \, \psi_h(\mathbf{x}_h,\tau_h) \,e^{-b\tau_h} \ . \label{eq: method of characteristics}
\end{equation}
Introducing the notation
$$
\bar{f}_{hk}^0 := \int_{\Omega_k(\mathbf{x}_h)}\int_{\rm V}\int_0^t f_{hk}(\cdot,\mathbf{x}_k-c\mathbf{v}_k\tau_h,\cdot,\cdot,\mathbf{v}_k,\tau_h=0,\tau_k-\tau_h) \diff\tau_k\diff\mathbf{v}_k\diff\mathbf{x}_k
$$
and substituting~\eqref{eq: method of characteristics} into~\eqref{eq: new densityhkbeta} gives
\begin{align}
p^{\beta_h}_h(\mathbf{x}_h,t,\mathbf{v}_h)&=\dfrac{1}{|{{\rm V}}|}\int_0^t\frac{\varphi_h(\mathbf{x}_h,\tau_h)}{\psi_h(\mathbf{x}_h,\tau_h)}\int_{\Omega_k(\mathbf{x}_h)}\int_{\rm V}\int_0^tf_{hk}\diff\tau_k\diff\mathbf{v}_k\diff\mathbf{x}_k\diff\tau_h\nonumber\\ &= \dfrac{1}{|{{\rm V}}|} \int_0^t\varphi_h(\mathbf{x}_h,\tau_h)e^{-b\tau_h}\bar{f}_{hk}^0(\mathbf{x}_h-c \, \mathbf{v}_h\tau_h,t-\tau_h,\mathbf{v}_h)\diff\tau_h\nonumber\\ & = \dfrac{1}{|{{\rm V}}|} \int_0^t\varphi_h(\mathbf{x}_h,t-s)e^{-(t-s)(b+c \, \mathbf{v}_h\cdot\nabla_{\mathbf{x}_h})}\bar{f}_{hk}^0(\mathbf{x}_h,s,\mathbf{v}_h)\diff s\ . \label{eq: almost there}
\end{align}
The last equality in~\eqref{eq: almost there} is obtained using the change of variables $s=t-\tau_h$ along with the following Taylor expansion
\begin{align}
    e^{-(t-s)c\mathbf{v}\cdot\nabla}f(\mathbf{x})&=\sum_{m=0}^\infty\frac{(-(t-s) \, c \, \mathbf{v}\cdot\nabla)^m}{m!}f(\mathbf{x})\nonumber\\ & =\sum_{m=0}^\infty\frac{1}{m!}(-(t-s) \, c \, \mathbf{v})^m\nabla^mf(\mathbf{x})=f(\mathbf{x}-(t-s) \, c \, \mathbf{v})\ .\nonumber
\end{align}
Hence, the Laplace transform in time of $p^{\beta_h}_h(\mathbf{x}_h,t,\mathbf{v}_h)$ is
\begin{equation}
    \hat{p}^{\beta_h}_h(\mathbf{x}_h,\lambda,\mathbf{v}_h) =\dfrac{1}{|{{\rm V}}|} \hat{\varphi}_h(\mathbf{x}_h,\lambda+b+c \, \mathbf{v}_h\cdot\nabla_{\mathbf{x}_h})\hat{\bar{f}}_{hk}^0(\mathbf{x}_h,\lambda,\mathbf{v}_h)\ .\label{eq: l transformed}
\end{equation}
Here, $\lambda$ is the Laplace variable, and $\hat{\varphi}_h$ and $\hat{\bar{f}}_{hk}^0$ are the Laplace transforms in time of the functions $\varphi_h$ and $\bar{f}_{hk}^0$. Moreover, substituting~\eqref{eq: method of characteristics} into~\eqref{eq: p densityhk} and computing the Laplace transform in time yields
$$
    \hat{p}_h(\mathbf{x}_h,\lambda,\mathbf{v}_h)=\dfrac{1}{|{{\rm V}}|} \hat{\psi}_h(\mathbf{x}_h,\lambda+b+c \, \mathbf{v}_h\cdot\nabla_{\mathbf{x}_h})\hat{\bar{f}}_{hk}^0(\mathbf{x}_h,\lambda,\mathbf{v}_h)\ ,
$$
with $\hat{\psi}_h$ being the Laplace transform of the function $\psi_h$. The latter equation gives
$$
\hat{\bar{f}}_{hk}^0(\mathbf{x}_h,\lambda,\mathbf{v}_h) = |{{\rm V}}| \, \dfrac{\hat{p}_h(\mathbf{x}_h,\lambda,\mathbf{v}_h)}{\hat{\psi}_h(\mathbf{x}_h,\lambda+b+c \, \mathbf{v}_h\cdot\nabla_{\mathbf{x}_h})}.
$$
Substituting such an expression for $\hat{\bar{f}}_{hk}^0$ into~\eqref{eq: l transformed} ones sees that~\eqref{eq: almost there} can be written as 
$$
  p^{\beta_h}_h(\mathbf{x}_h,t,\mathbf{v}_h)=\mathcal{B}[p_h](\mathbf{x}_h,t,\mathbf{v}_h)
$$
with the integral operator $\mathcal{B}$ being defined according to~\eqref{def:operatorB}. 

\bibliographystyle{plain}
\bibliography{LevyBrownian}

\begin{thebibliography}{10}

\bibitem{albrecht1977phagokinetic}
Guenter Albrecht-Buehler.
\newblock The phagokinetic tracks of 3{T}3 cells.
\newblock {\em Cell}, 11(2):395--404, 1977.

\bibitem{alt1980biased}
Wolgang Alt.
\newblock Biased random walk models for chemotaxis and related diffusion
  approximations.
\newblock {\em Journal of Mathematical Biology}, 9(2):147--177, 1980.

\bibitem{azizi2018single}
Elham Azizi, Ambrose~J Carr, George Plitas, Andrew~E Cornish, Catherine
  Konopacki, Sandhya Prabhakaran, Juozas Nainys, Kenmin Wu, Vaidotas
  Kiseliovas, Manu Setty, et~al.
\newblock Single-cell map of diverse immune phenotypes in the breast tumor
  microenvironment.
\newblock {\em Cell}, 174(5):1293--1308, 2018.

\bibitem{bartumeus2003helical}
Frederic Bartumeus, Francesc Peters, Salvador Pueyo, Celia Marras{\'e}, and
  Jordi Catalan.
\newblock Helical {L}{\'e}vy walks: adjusting searching statistics to resource
  availability in microzooplankton.
\newblock {\em Proceedings of the National Academy of Sciences},
  100(22):12771--12775, 2003.

\bibitem{boissonnas2007vivo}
Alexandre Boissonnas, Luc Fetler, Ingrid~S Zeelenberg, St{\'e}phanie Hugues,
  and Sebastian Amigorena.
\newblock In vivo imaging of cytotoxic {T} cell infiltration and elimination of
  a solid tumor.
\newblock {\em The Journal of Experimental Medicine}, 204(2):345--356, 2007.

\bibitem{bousso2008t}
Philippe Bousso.
\newblock T-cell activation by dendritic cells in the lymph node: lessons from
  the movies.
\newblock {\em Nature Reviews Immunology}, 8(9):675--684, 2008.

\bibitem{cercignani2013mathematical}
Carlo Cercignani, Reinhard Illner, and Mario Pulvirenti.
\newblock {\em The mathematical theory of dilute gases}, volume 106.
\newblock Springer Science \& Business Media, 2013.

\bibitem{de2014superdiffusion}
Monique de~Jager, Frederic Bartumeus, Andrea K{\"o}lzsch, Franz~J Weissing,
  Geerten~M Hengeveld, Bart~A Nolet, Peter~MJ Herman, and Johan van~de Koppel.
\newblock How superdiffusion gets arrested: ecological encounters explain shift
  from {L}{\'e}vy to {B}rownian movement.
\newblock {\em Proceedings of the Royal Society B: Biological Sciences},
  281(1774):20132605, 2014.

\bibitem{de2007patch}
HJ~De~Knegt, GM~Hengeveld, F~Van~Langevelde, WF~De~Boer, and KP~Kirkman.
\newblock Patch density determines movement patterns and foraging efficiency of
  large herbivores.
\newblock {\em Behavioral Ecology}, 18(6):1065--1072, 2007.

\bibitem{engelhardt2012marginating}
John~J Engelhardt, Bijan Boldajipour, Peter Beemiller, Priya Pandurangi,
  Caitlin Sorensen, Zena Werb, Mikala Egeblad, and Matthew~F Krummel.
\newblock Marginating dendritic cells of the tumor microenvironment
  cross-present tumor antigens and stably engage tumor-specific {T} cells.
\newblock {\em Cancer Cell}, 21(3):402--417, 2012.

\bibitem{estrada2020interacting}
Gissell Estrada-Rodriguez and Heiko Gimperlein.
\newblock Interacting particles with {L}{\'e}vy strategies: limits of transport
  equations for swarm robotic systems.
\newblock {\em SIAM Journal on Applied Mathematics}, 80(1):476--498, 2020.

\bibitem{estrada2018fractional}
Gissell Estrada-Rodriguez, Heiko Gimperlein, and Kevin~J Painter.
\newblock Fractional {P}atlak--{K}eller--{S}egel equations for chemotactic
  superdiffusion.
\newblock {\em SIAM Journal on Applied Mathematics}, 78(2):1155--1173, 2018.

\bibitem{franz2016hard}
Benjamin Franz, Jake~P Taylor-King, Christian Yates, and Radek Erban.
\newblock Hard-sphere interactions in velocity-jump models.
\newblock {\em Physical Review E}, 94(1):012129, 2016.

\bibitem{gardner2016dendritic}
Alycia Gardner and Brian Ruffell.
\newblock Dendritic cells and cancer immunity.
\newblock {\em Trends in Immunology}, 37(12):855--865, 2016.

\bibitem{humphries2010environmental}
Nicolas~E Humphries, Nuno Queiroz, Jennifer~RM Dyer, Nicolas~G Pade, Michael~K
  Musyl, Kurt~M Schaefer, Daniel~W Fuller, Juerg~M Brunnschweiler, Thomas~K
  Doyle, Jonathan~DR Houghton, et~al.
\newblock Environmental context explains {L}{\'e}vy and {B}rownian movement
  patterns of marine predators.
\newblock {\em Nature}, 465(7301):1066--1069, 2010.

\bibitem{kennard1938kinetic}
Earle~H Kennard et~al.
\newblock {\em Kinetic theory of gases}, volume 287.
\newblock McGraw-hill New York, 1938.

\bibitem{krummel2016t}
Matthew~F Krummel, Frederic Bartumeus, and Audrey G{\'e}rard.
\newblock {T} cell migration, search strategies and mechanisms.
\newblock {\em Nature Reviews Immunology}, 16(3):193, 2016.

\bibitem{lober2015collisions}
Jakob L{\"o}ber, Falko Ziebert, and Igor~S Aranson.
\newblock Collisions of deformable cells lead to collective migration.
\newblock {\em Scientific Reports}, 5(1):1--7, 2015.

\bibitem{macfarlane2019stochastic}
Fiona~R Macfarlane, Mark~AJ Chaplain, and Tommaso Lorenzi.
\newblock A stochastic individual-based model to explore the role of spatial
  interactions and antigen recognition in the immune response against solid
  tumours.
\newblock {\em Journal of Theoretical Biology}, 480:43--55, 2019.

\bibitem{macfarlane2018modelling}
Fiona~R Macfarlane, Tommaso Lorenzi, and Mark~AJ Chaplain.
\newblock Modelling the immune response to cancer: an individual-based approach
  accounting for the difference in movement between inactive and activated {T}
  cells.
\newblock {\em Bulletin of Mathematical Biology}, 80(6):1539--1562, 2018.

\bibitem{nolet2002search}
Bart~A Nolet and Wolf~M Mooij.
\newblock Search paths of swans foraging on spatially autocorrelated tubers.
\newblock {\em Journal of Animal Ecology}, pages 451--462, 2002.

\bibitem{othmer1988models}
Hans~G Othmer, Steven~R Dunbar, and Wolfgang Alt.
\newblock Models of dispersal in biological systems.
\newblock {\em Journal of Mathematical Biology}, 26(3):263--298, 1988.

\bibitem{othmer2000diffusion}
Hans~G Othmer and Thomas Hillen.
\newblock The diffusion limit of transport equations derived from velocity-jump
  processes.
\newblock {\em SIAM Journal on Applied Mathematics}, 61(3):751--775, 2000.

\bibitem{othmer2012experimental}
Hans~G Othmer, Philip~K Maini, and James~D Murray.
\newblock {\em Experimental and theoretical advances in biological pattern
  formation}, volume 259.
\newblock Springer Science \& Business Media, 2012.

\bibitem{perthame2018fractional}
Beno{\^\i}t Perthame, Weiran Sun, and Min Tang.
\newblock The fractional diffusion limit of a kinetic model with biochemical
  pathway.
\newblock {\em Zeitschrift f{\"u}r angewandte Mathematik und Physik}, 69(3):67,
  2018.

\bibitem{rothoeft2006structure}
Tobias Rothoeft, Sandra Balkow, Mathias Krummen, Stefan Beissert, Georg Varga,
  Karin Loser, Pia Oberbanscheidt, Frank van~den Boom, and Stephan Grabbe.
\newblock Structure and duration of contact between dendritic cells and {T}
  cells are controlled by {T} cell activation state.
\newblock {\em European Journal of Immunology}, 36(12):3105--3117, 2006.

\bibitem{villani2002review}
C{\'e}dric Villani.
\newblock A review of mathematical topics in collisional kinetic theory.
\newblock {\em Handbook of mathematical fluid dynamics}, 1(71-305):3--8, 2002.

\bibitem{waldman2020guide}
Alex~D Waldman, Jill~M Fritz, and Michael~J Lenardo.
\newblock A guide to cancer immunotherapy: from {T} cell basic science to
  clinical practice.
\newblock {\em Nature Reviews Immunology}, pages 1--18, 2020.

\bibitem{wculek2019dendritic}
Stefanie~K Wculek, Francisco~J Cueto, Adriana~M Mujal, Ignacio Melero,
  Matthew~F Krummel, and David Sancho.
\newblock Dendritic cells in cancer immunology and immunotherapy.
\newblock {\em Nature Reviews Immunology}, pages 1--18, 2019.

\bibitem{wherry2015molecular}
E~John Wherry and Makoto Kurachi.
\newblock Molecular and cellular insights into {T} cell exhaustion.
\newblock {\em Nature Reviews Immunology}, 15(8):486--499, 2015.

\bibitem{witt2005directed}
Colleen~M Witt, Subhadip Raychaudhuri, Brian Schaefer, Arup~K Chakraborty, and
  Ellen~A Robey.
\newblock Directed migration of positively selected thymocytes visualized in
  real time.
\newblock {\em PLOS Biolology}, 3(6):e160, 2005.

\bibitem{worbs2007ccr7}
Tim Worbs, Thorsten~R Mempel, Jasmin B{\"o}lter, Ulrich~H von Andrian, and
  Reinhold F{\"o}rster.
\newblock {CCR}7 ligands stimulate the intranodal motility of {T} lymphocytes
  in vivo.
\newblock {\em The Journal of Experimental Medicine}, 204(3):489--495, 2007.

\end{thebibliography}

\end{document}